\documentclass[preprint]{ptephy_v1}

\preprintnumber{arXive:2103.00101 }
\usepackage{bm}
\usepackage{wasysym}
\usepackage{graphics} 
\usepackage{subfig} 

\begin{document}

\title{Development of a cavity with photonic crystal structure for axion searches}


\author[1,2]{Y.~Kishimoto}
\affil{ Research Center for Neutrino Science, Tohoku University, Sendai 980-8578, Japan \email{kisimoto@awa.tohoku.ac.jp}}
\affil[2]{ Kavli Institute for the Physics and Mathematics of the Universe (WPI), the University of Tokyo, Kashiwa, Chiba, 277-8582, Japan}

\author[1]{Y.~Suzuki}
\author[3]{I.~Ogawa}
\affil[3]{Faculty of Engineering, University of Fukui, Fukui 910-8507, Japan}
\author[4]{Y.~Mori} 
\affil[4]{Graduate School of Engineering, University of Fukui, Fukui 910-8507, Japan}

\author[5,2,6]{M.~Yamashita}
\affil[5]{Kamioka Observatory, Institute for Cosmic Ray Research, University of Tokyo, Kamioka, Gifu 506-1205, Japan}

\affil[6]{Present address: Institute for Space-Earth Environmental Research, Nagoya University, Nagoya Aichi, 464-8601, Japan}


\begin{abstract}%
Two cavities in different sizes with the photonic crystal structure have been developed for axion searches.
In the cavities, the dispersion relation in the photonic crystal is utilized, and so they are named to "DRiPC cavities".  
The size of the smaller one is 100 mm $\times$ 100 mm $\times$ 10 mm, 
where 16 cylindrical metal poles with a diameter of 4 mm are introduced in a 4$\times$4 grid at 20 mm intervals.
In this study, the grid interval in $x$ direction in the small size cavity, $L_x$, was changed to investigate resonance frequency, $Q$-value, and electric field profile at each $L_x$.
The lowest three frequencies have been compared with the ones simulated by the finite element method to be found in excellent agreement.
The lowest frequency mode could be tuned from 5.10 GHz ($L_x$ = 25.0 mm) to 6.72 GHz (13.9 mm), centering on 5.87 GHz at $L_x$ = 20 mm.
This wide range tunability, 27.7\%, was suitable 
for a search with a modest $Q$-value.
By examining the electric field distributions with the bead pull method, the lowest frequency mode at $L_x=16.0$ -- 25.0 mm were ${\rm TM_{010}}$-like.
This mode was also obtained in a larger size cavity 
(180 mm $\times$ 180 mm $\times$ 20 mm $\times$ 2)
with the same photonic crystal structure.
These results led us to conclude a DRiPC cavity has the noble features for future axion search experiments.
\end{abstract}

\subjectindex{H2}

\maketitle

\section{Introduction} \label{sec:introduction}

Many astronomical observations have confirmed the existence of dark matter.
However, the nature of the dark matter remains unknown, and many unknown elementary particles are being actively discussed as candidates, such as WIMPs and axions.
The latter, axion,  is an elementary particle proposed to solve the $CP$ problem in the strong interaction and is one of the promising candidates for the dark matter.
One of the characteristics of the dark matter axion is that it has a very light mass, from 1 $\mu$eV to 10 meV.
Such light dark matter is called weakly interacting slim particles (WISPs),
and it includes Axion-Like Particles (ALPs), hidden photons, and so forth besides axion.
Various methods have been proposed, developed, and implemented in the searches for these WISPs, corresponding to the wide mass range, 
from the pioneering work by RBF \cite{bib:RBF} 
to the currently running programs,
such as ADMX \cite{bib:admx1, bib:admx2, bib:admx3, bib:admx4},
HAYSTAC \cite{bib:haystac1, bib:haystac2},
ORGAN \cite{bib:oregon},
experiments in CAPP \cite{bib:capp1, bib:capp2},
QUAX-a$\gamma$ \cite{bib:quax},
KLASH (proposal) \cite{bib:klash},
and
RADES (under construction) \cite{bib:rades},
as well as new ideas
by MADMAX \cite{bib:madmax}, 
CASPer \cite{bib:casper}, and so on.

In many of these experiments,
a method of measuring axion (or ALP)-converted photons generated in a strong magnetic field with a resonant cavity, called "halo-scope" \cite{bib:haloscopeSikivie1, bib:haloscopeSikivie2, bib:haloscopeKrauss},
is utilized.
In the halo-scope, signal power, $P$, is given by 
\begin{equation}\label{eqn:axionPower}
	P = \kappa g_{a\gamma} (\rho / m_a) B_0^2 V G Q, 
\end{equation}	
where $\kappa$ is a coupling to the cavity, $g_{a\gamma}$, $\rho$, $m_a$ 
are axion-photon coupling, dark matter energy density around the Solar system and axion mass, respectively.
$B_0$ is external magnetic field, and $V$ and $Q$ are volume and $Q$-value of the cavity.
$G$ is geometric factor  ($G$-factor), given by 
\begin{equation}\label{eqn:gFactor}
	G=\frac{(\int_V dv {\bm E}_{\rm cav}  \cdot {\bm B}_0)^2 }{VB_0^2 \int_V dv {\bm E}_{\rm cav} \cdot \bm{E}_{\rm cav} } ,
\end{equation}	
where ${\bm E}_{\rm cav}$ is the resonant electric-field in the cavity.
It is clear that an experiment using a large-volume and high-$Q$ cavity in a strong magnetic field is effective.
In Eqn. \ref{eqn:gFactor}, however, 
it is required 
that ${\bm E}_{\rm cav}$  and ${\bm B}_0$ are parallel, 
while if they are anti-parallel, that part negatively contributes to the integral.
Therefore, the mode without node, such as ${\rm TM_{010}}$, is optimal.
However, taking a cylindrical cavity whose radius is $r$ as a typical example, 
the resonance frequency of ${\rm TM_{010}}$ mode, $f$, is given by $f=\frac{2.405c}{2\pi r}$.
At high frequencies, therefore, it is necessary for $r$ to be small, so $V$ is inevitably limited.
This means the power of the axion-converted photon gets smaller for a higher frequency, making experiments difficult.
To overcome the issue, intense R\&D activities to enlarge cavity volume \cite{bib:capp2, bib:multiCAAP1, bib:multiCAAP2, bib:multiRades},
to utilize higher $Q$-cavity \cite{bib:scCavityNbTi, bib:scCavityYBCO},
and to exploit higher-order modes \cite{bib:quaxHighOder, bib:highOderMode1, bib:highOderMode2}
have been advanced.
As related to the volume enhancement, usage of a photonic crystal structure was briefly mentioned in the literature \cite{bib:admxPhotonic},
but publications on detailed experimental results have not been found to date.
In terms of  the photonic crystal cavity, 
many studies have been actively conducted to utilize the photonic band gap  (PBG) generated by its crystal structure,
in particular, the RF cavities for accelerators
\cite{bib:accPBGCavity}
and also for  axion researches \cite{bib:admxPBGCavity, bib:haystacPBGCavity, bib:quaxPBGCavity},
as well. 

In this paper, we focus on another essential aspect in the photonic crystal structure: 
the dispersion relation is determined by the crystal length \cite{bib:photonicCavityDispersion}, that is, resonance frequencies are
independent from the cavity size.
It therefore can open the possibility of a larger resonant cavity 
in a higher frequency regime. 
We call a large cavity based on this feature {\it Dispersion Relation in Photonic Crystal}-cavity  (DRiPC-cavity). 
We report the development of  two DRiPC cavities in the range of 5 to 6 GHz, 
the relation between resonance frequency and crystal period, 
and profiles of the electric field.
In Sec. \ref{sec:design} designs of the cavities are described,  and measurements and results are followed in Sec. \ref{sec:measAndResults}.
We summarize in Sec. \ref{sec:conclusion}.

\section{Design of the DRiPC cavity}\label{sec:design}

\begin{figure}[!h]
\begin{center}
\centering
	\subfloat[][Base design of the DRiPC cavity]{\includegraphics[width=2.5in, angle=90]{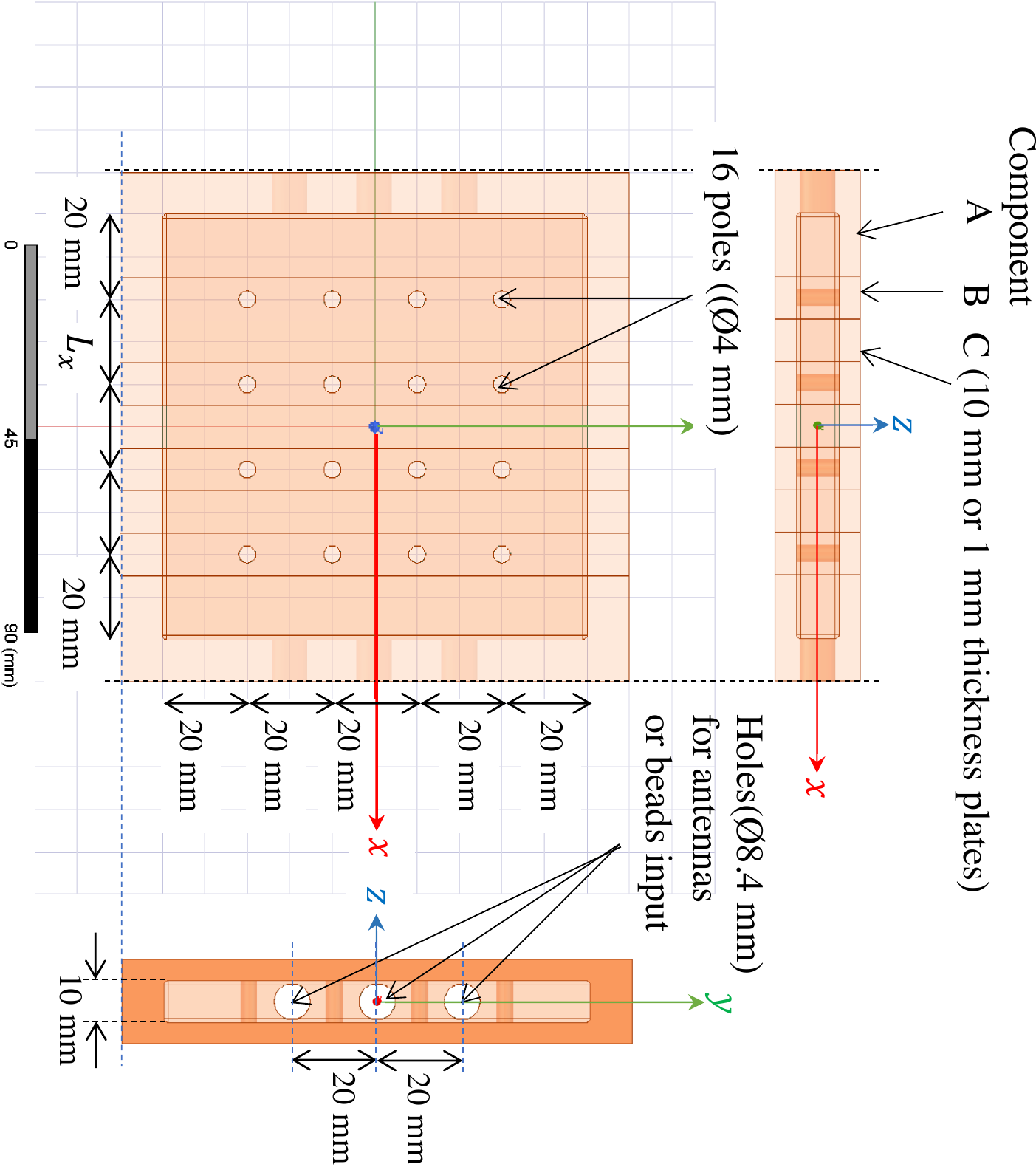}} \label{subfig:Design} 
	\qquad
	\subfloat[][Electric field vector on $xy$ plane, ${\bm E}(x, y, z=0)$] {\includegraphics[width=2.5in, angle=00, clip]{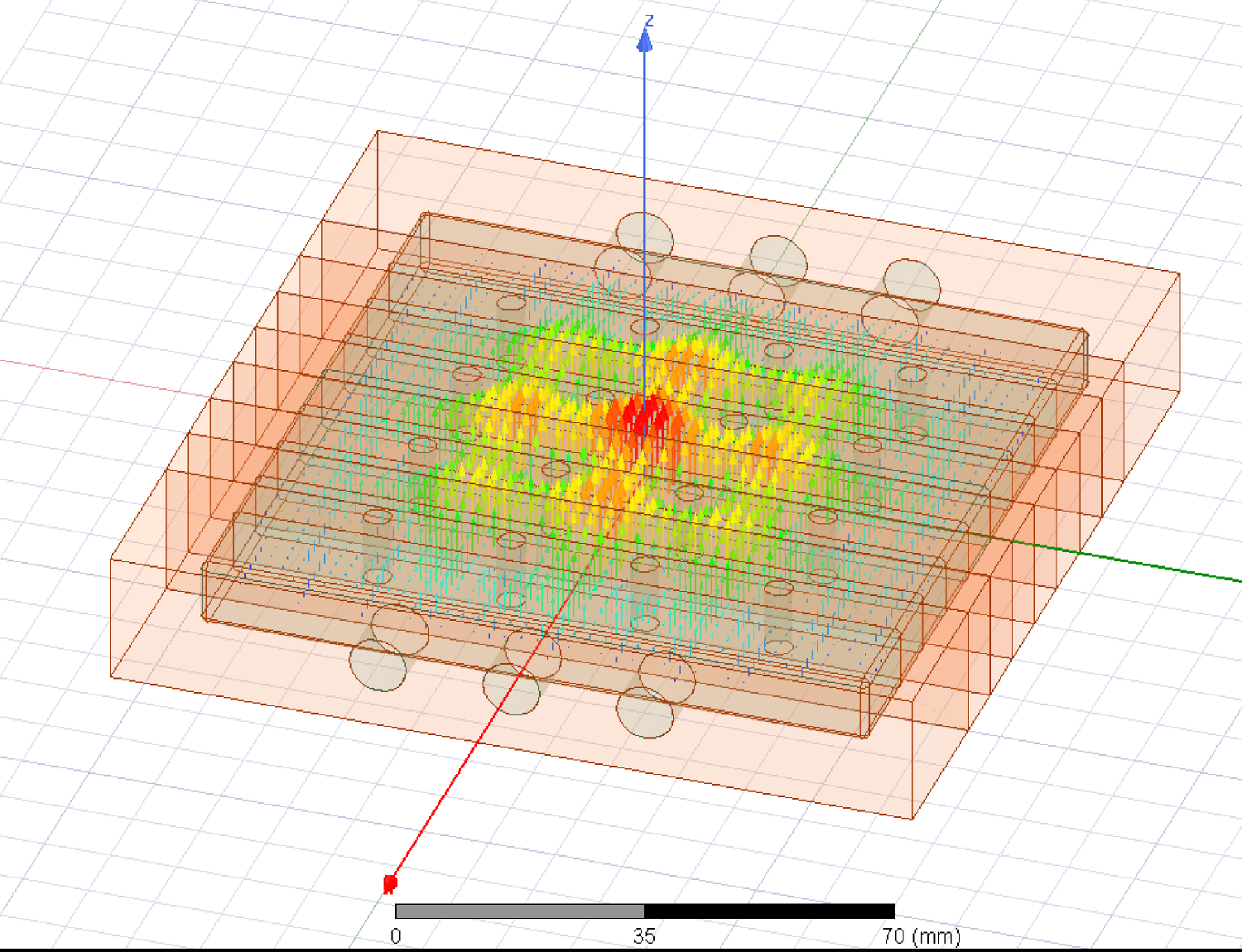}} \label{subfig:Vector}
 	\captionsetup{format=plain }
	\caption[]{ (a) The base design of the cavity in a three view drawing. The grid spacing in $x$ direction, $L_x$, is changed to tune the resonance frequency. 
	Six holes are provided at both $x$-ends for access into the inside the cavity, such as antennas.
	(b) Electric field vector on $xy$ plane, ${\bm E}(x, y, z=0)$, in the case of $L_x$=20 mm. All of the field direction is parallel to $z$ axis.}
	
\label{fig:designAndVector}
\end{center}
\end{figure}

\begin{figure}[!h]
\begin{center}
\centering
\subfloat[] [$L_x$=14 mm]{\includegraphics[width=1.5in, angle=270, clip]{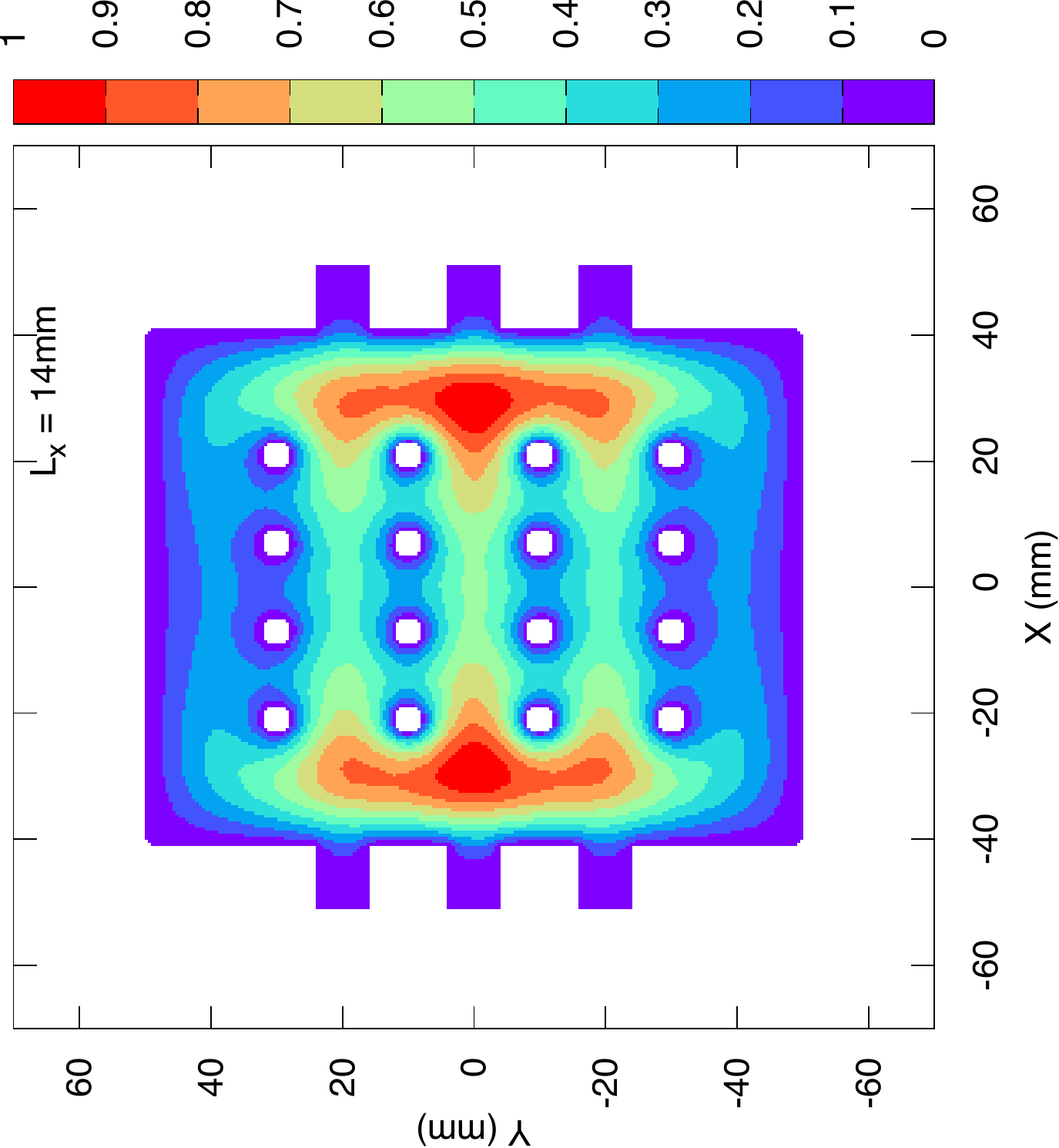}} \label{fig:Lx14mm}
\subfloat[] [$L_x$=15 mm]{\includegraphics[width=1.5in, angle=270, clip]{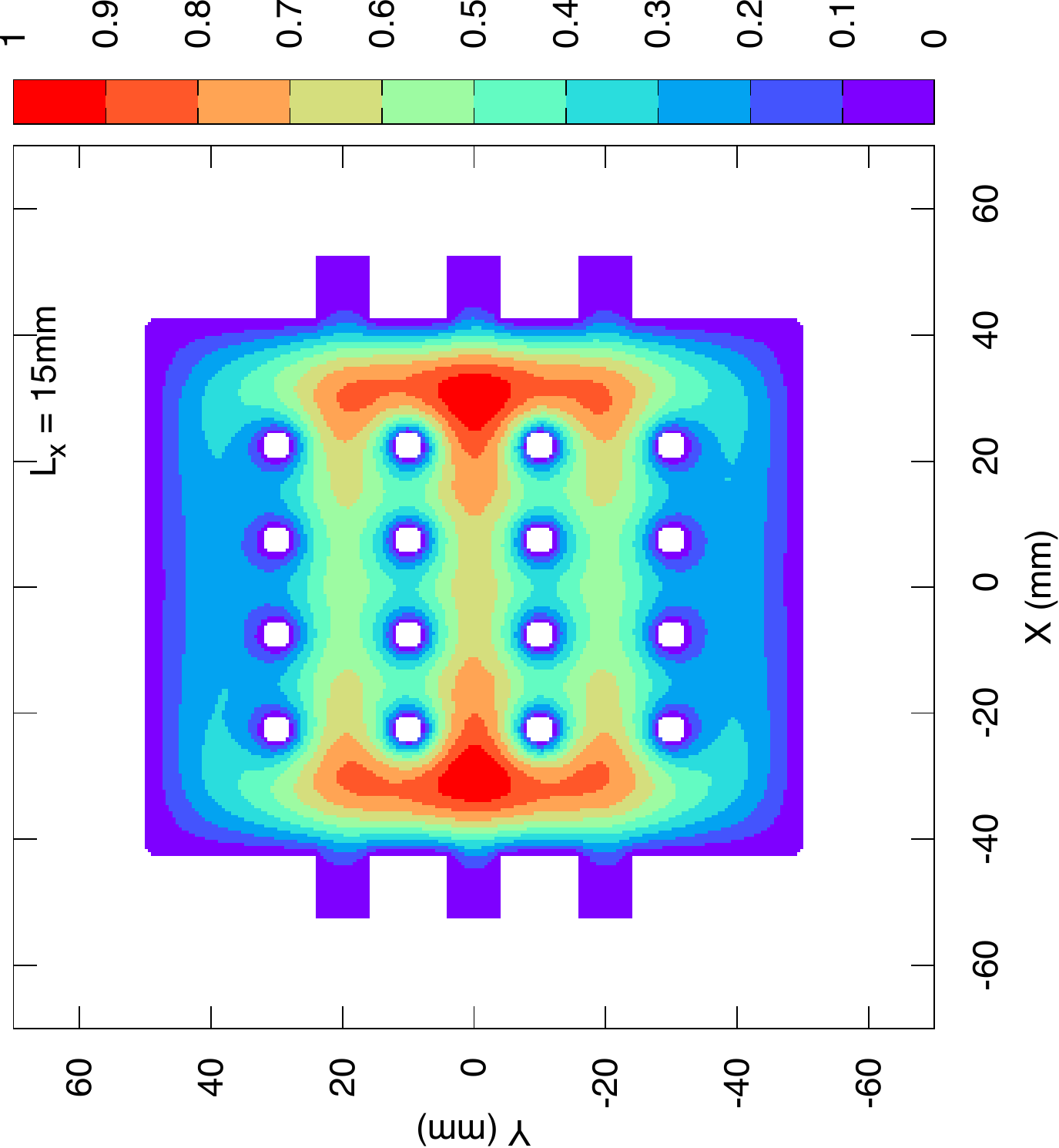}} \label{fig:Lx15mm}
\subfloat[] [$L_x$=16 mm]{\includegraphics[width=1.5in, angle=270, clip]{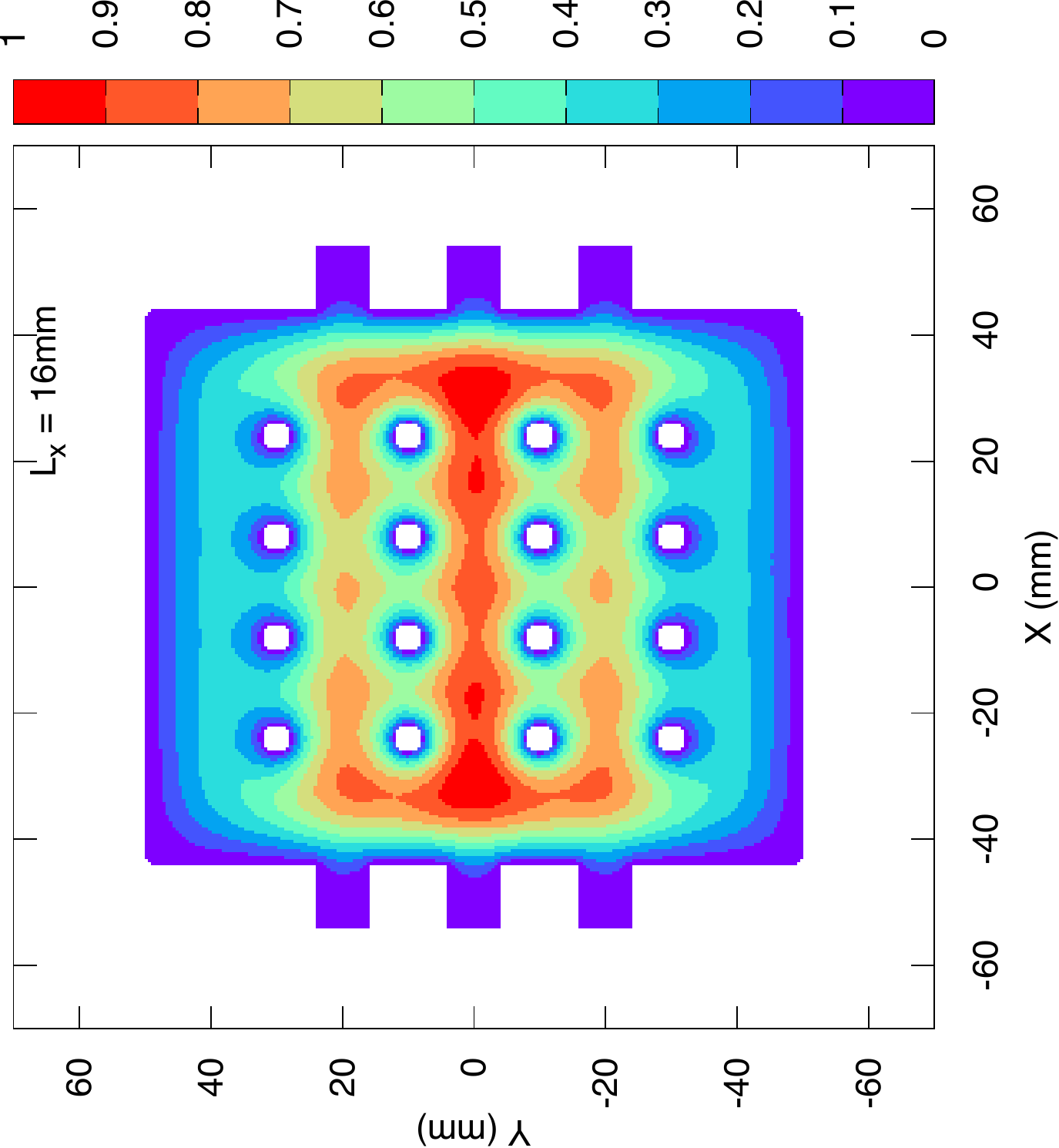}} \label{fig:Lx16mm}

\subfloat[] [$L_x$=17 mm]{\includegraphics[width=1.5in, angle=270, clip]{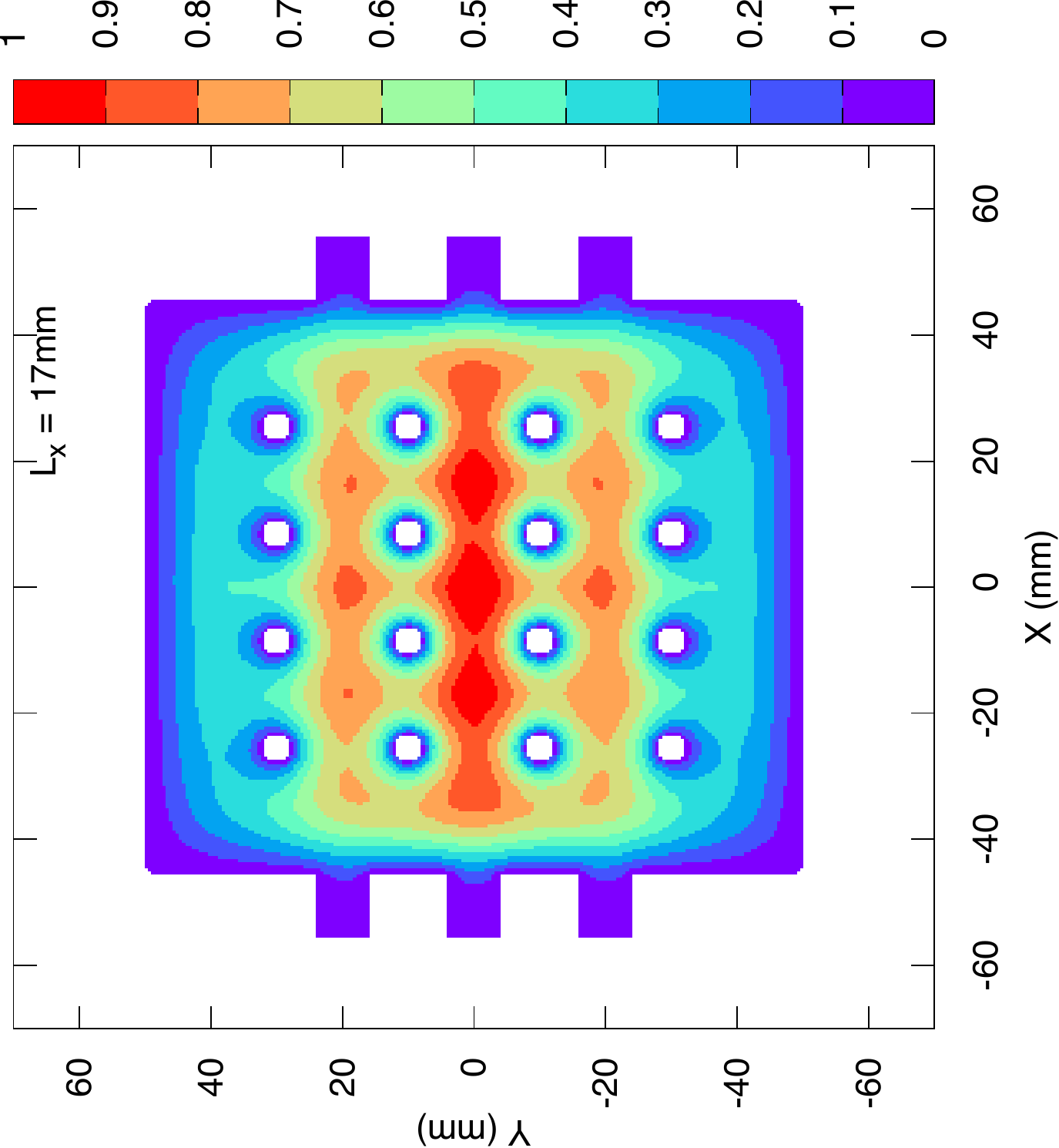}} \label{fig:Lx17mm}
\subfloat[] [$L_x$=18 mm]{\includegraphics[width=1.5in, angle=270, clip]{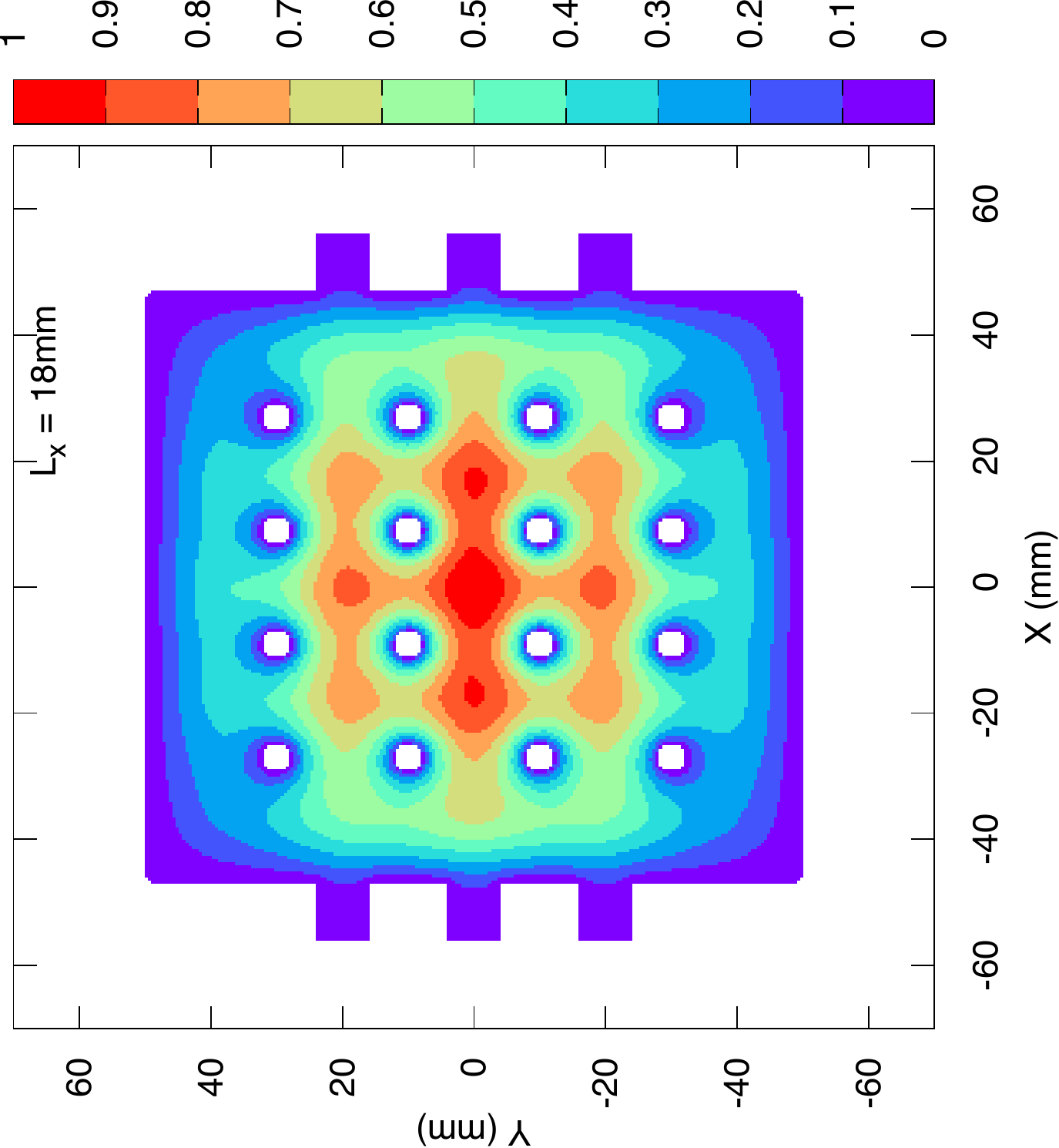}} \label{fig:Lx18mm}
\subfloat[] [$L_x$=19 mm]{\includegraphics[width=1.5in, angle=270, clip]{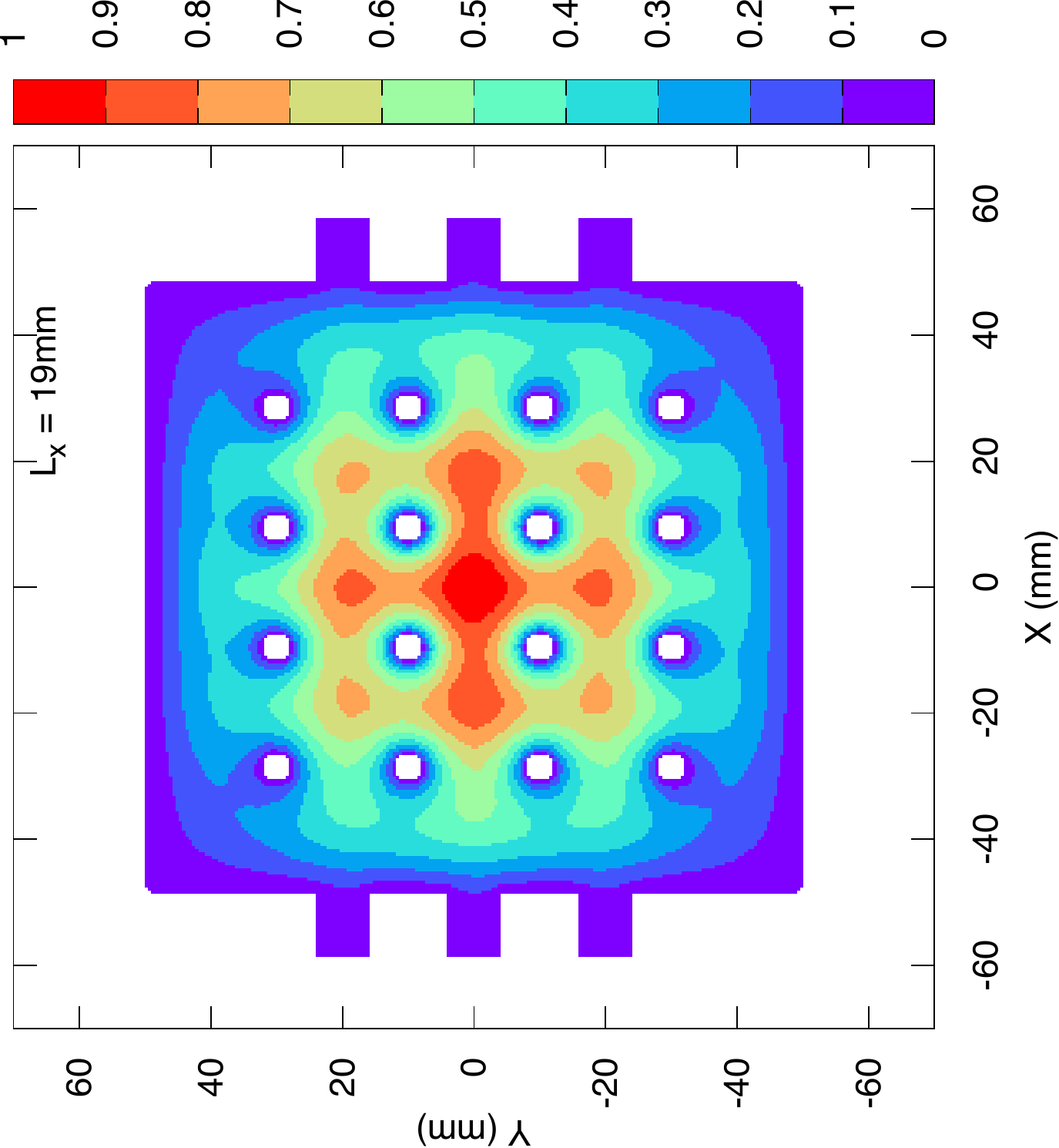}} \label{fig:Lx19mm}

\subfloat[] [$L_x$=20 mm]{\includegraphics[width=1.5in, angle=270, clip]{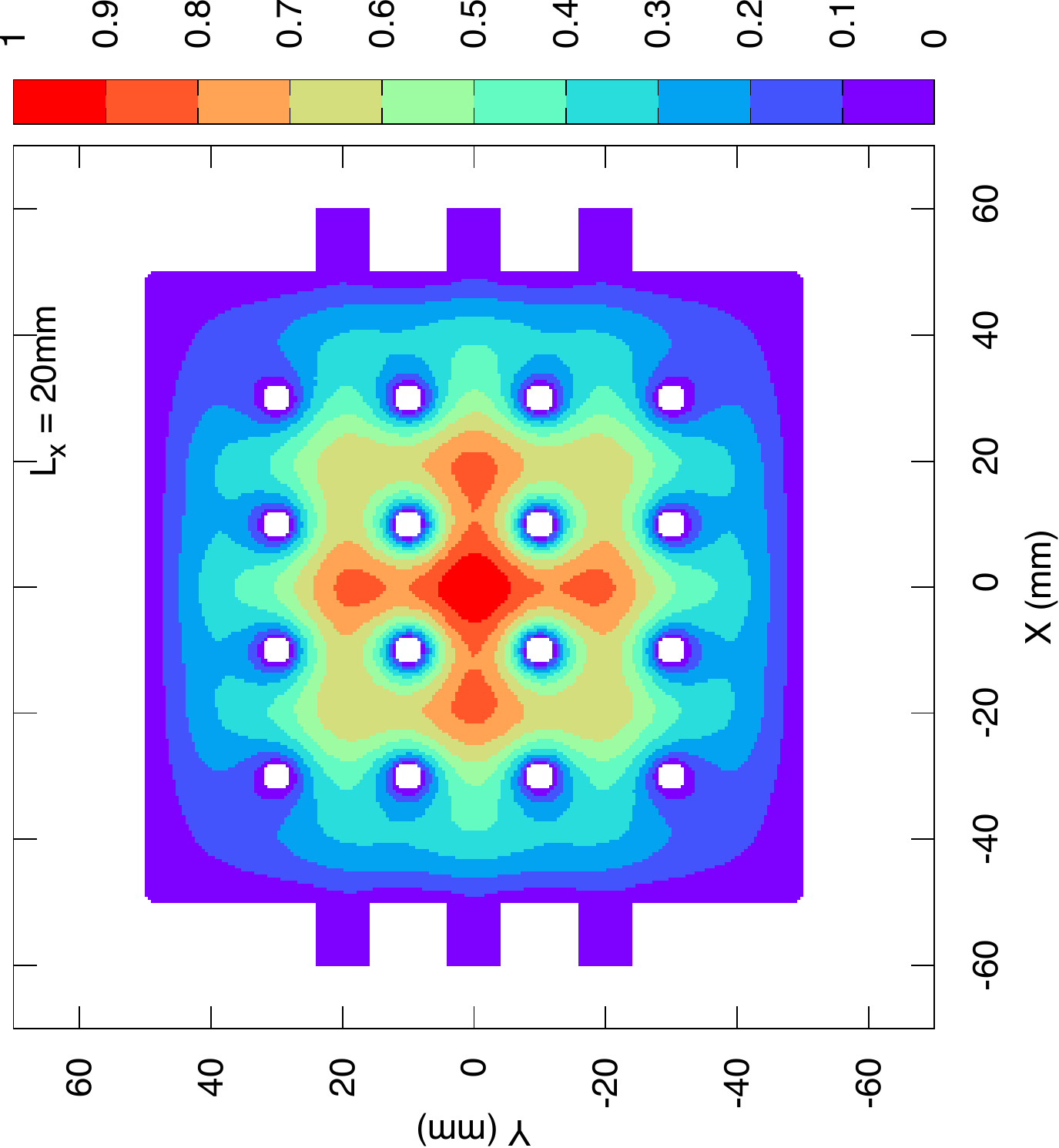}} \label{fig:Lx20mm}
\subfloat[] [$L_x$=21 mm]{\includegraphics[width=1.5in, angle=270, clip]{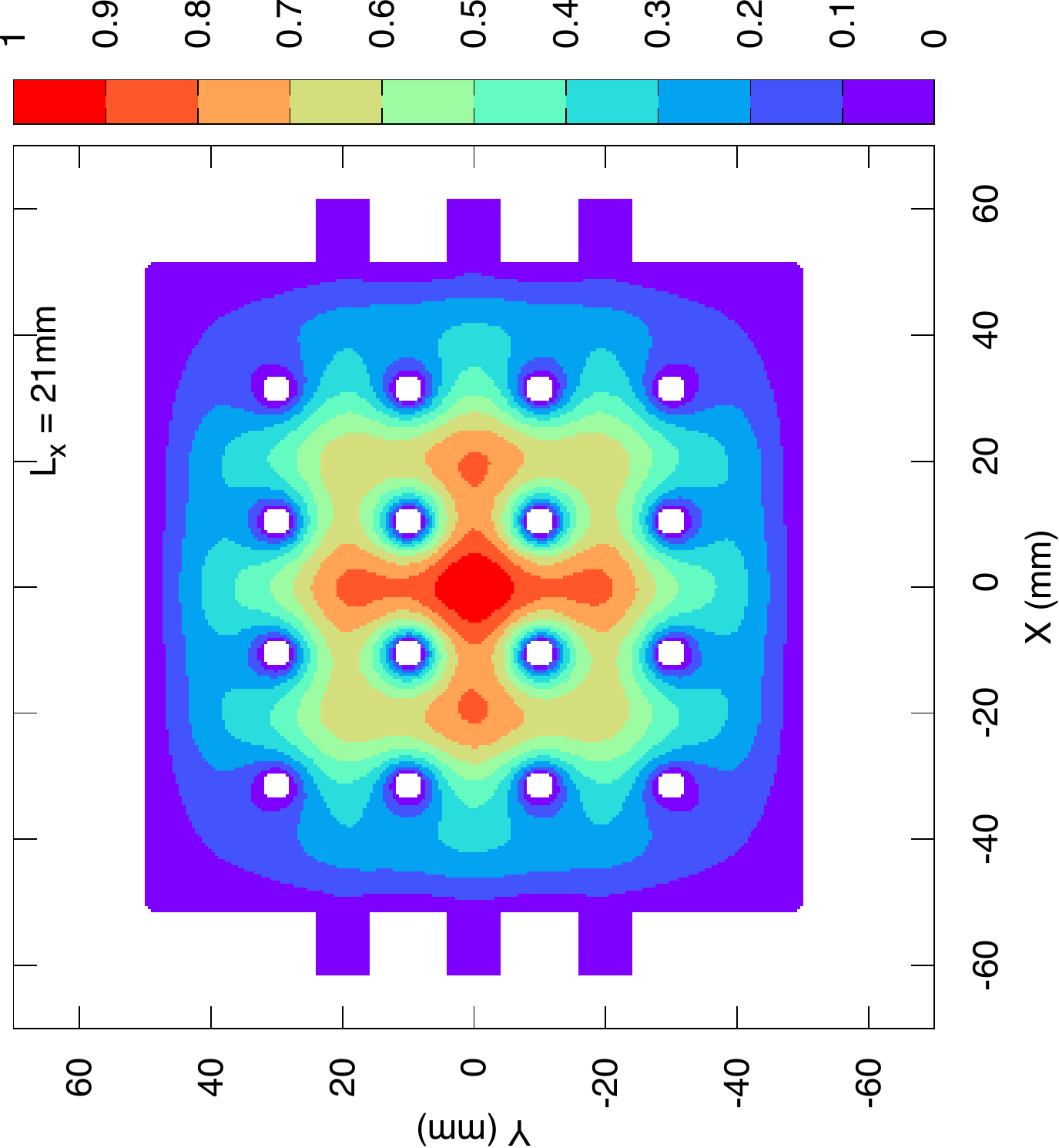}} \label{fig:Lx21mm}
\subfloat[] [$L_x$=22 mm]{\i\includegraphics[width=1.5in, angle=270, clip]{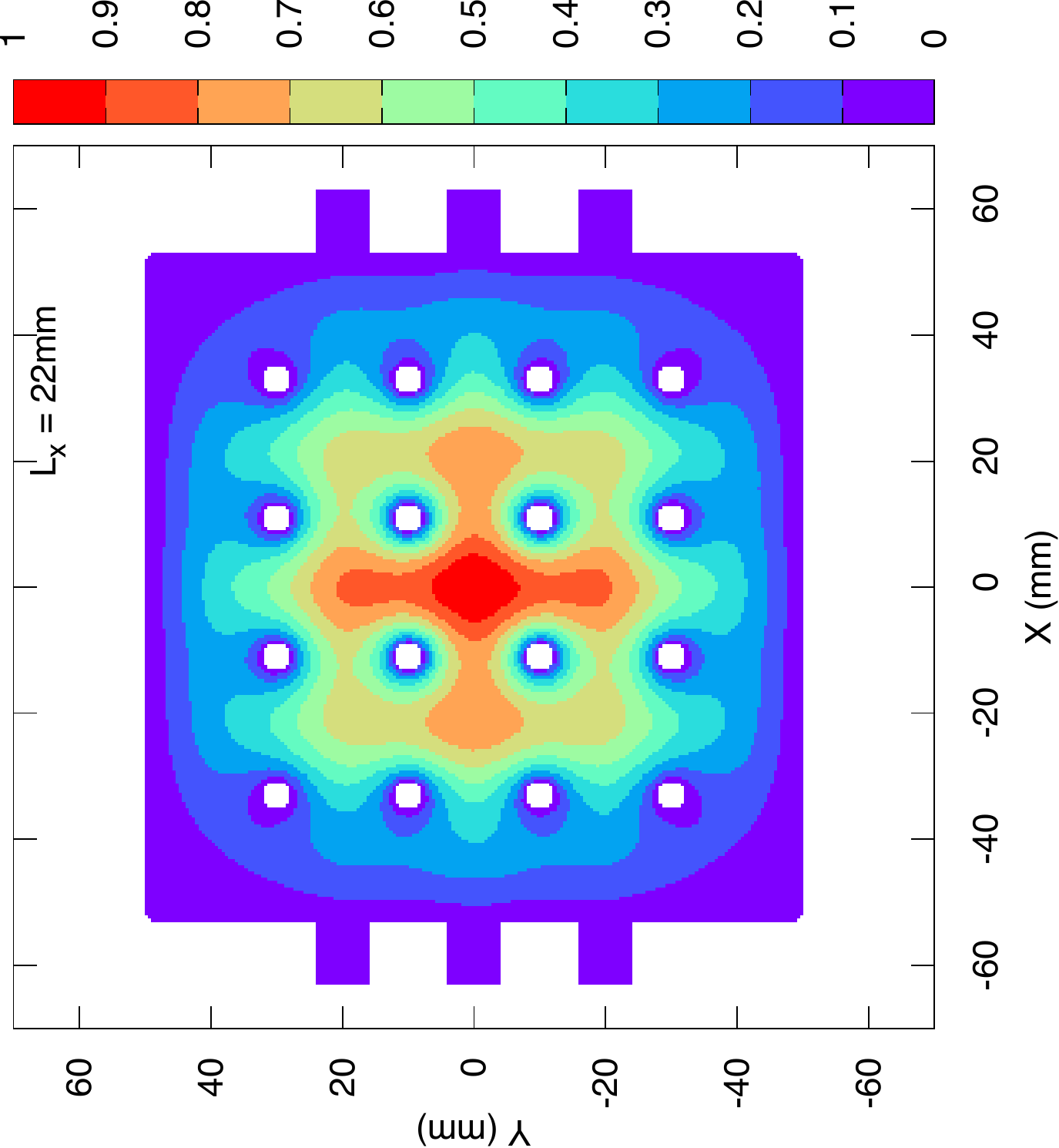}} \label{fig:Lx22mm}

\subfloat[] [$L_x$=23 mm]{\includegraphics[width=1.5in, angle=270, clip]{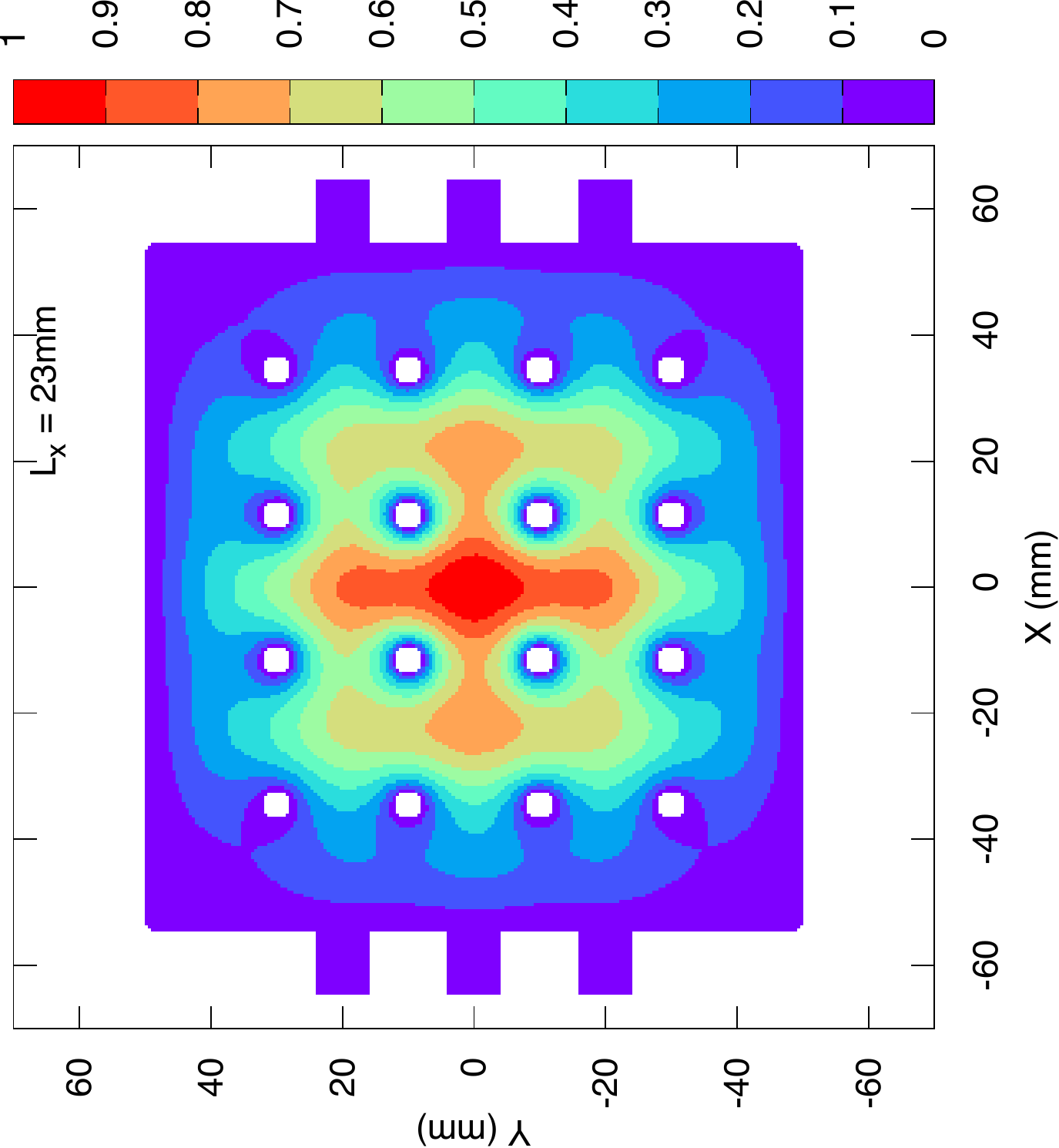}} \label{fig:Lx23mm}
\subfloat[] [$L_x$=24 mm]{\includegraphics[width=1.5in, angle=270, clip]{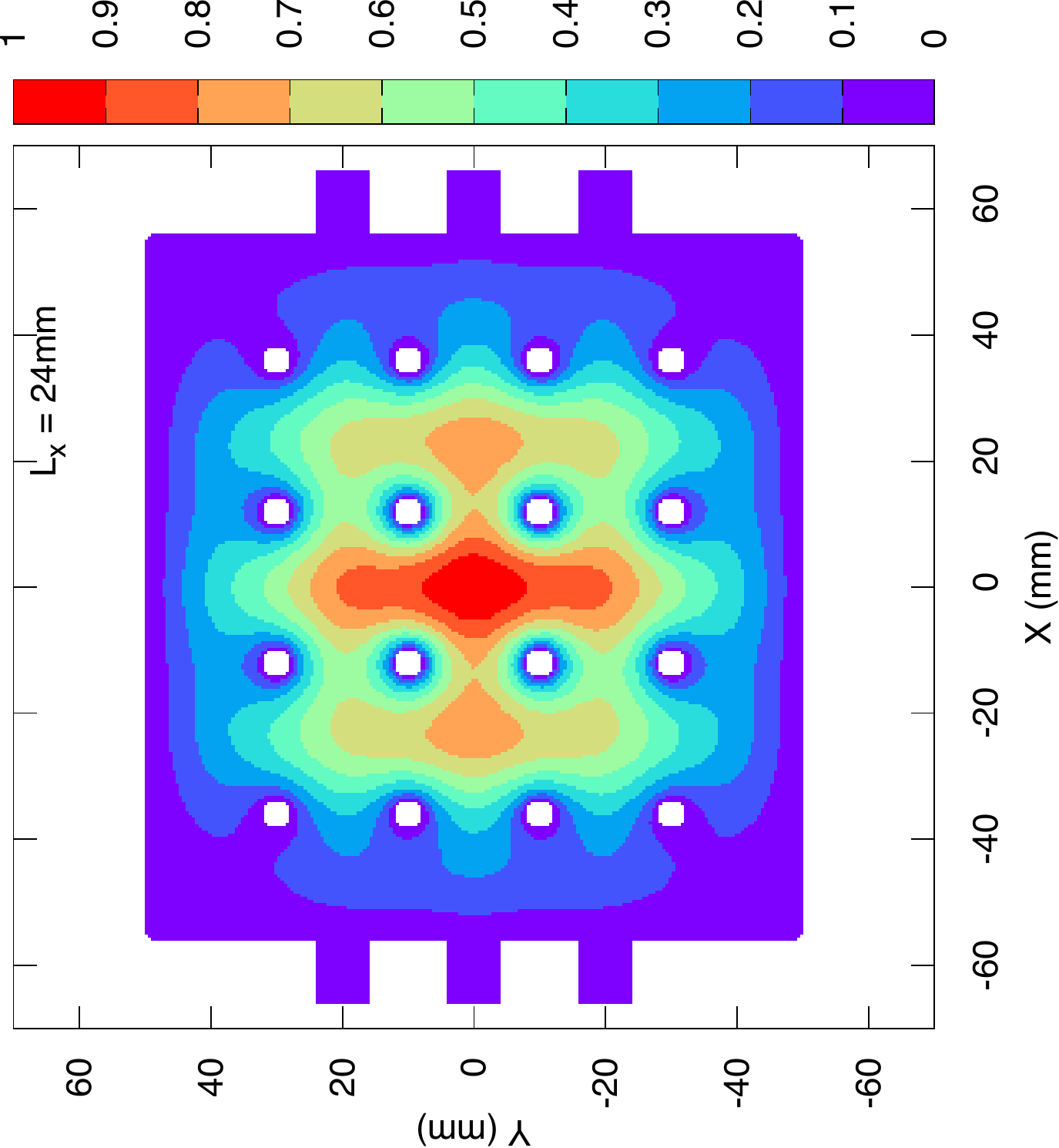}} \label{fig:Lx24mm}
\subfloat[] [$L_x$=25 mm]{\includegraphics[width=1.5in, angle=270, clip]{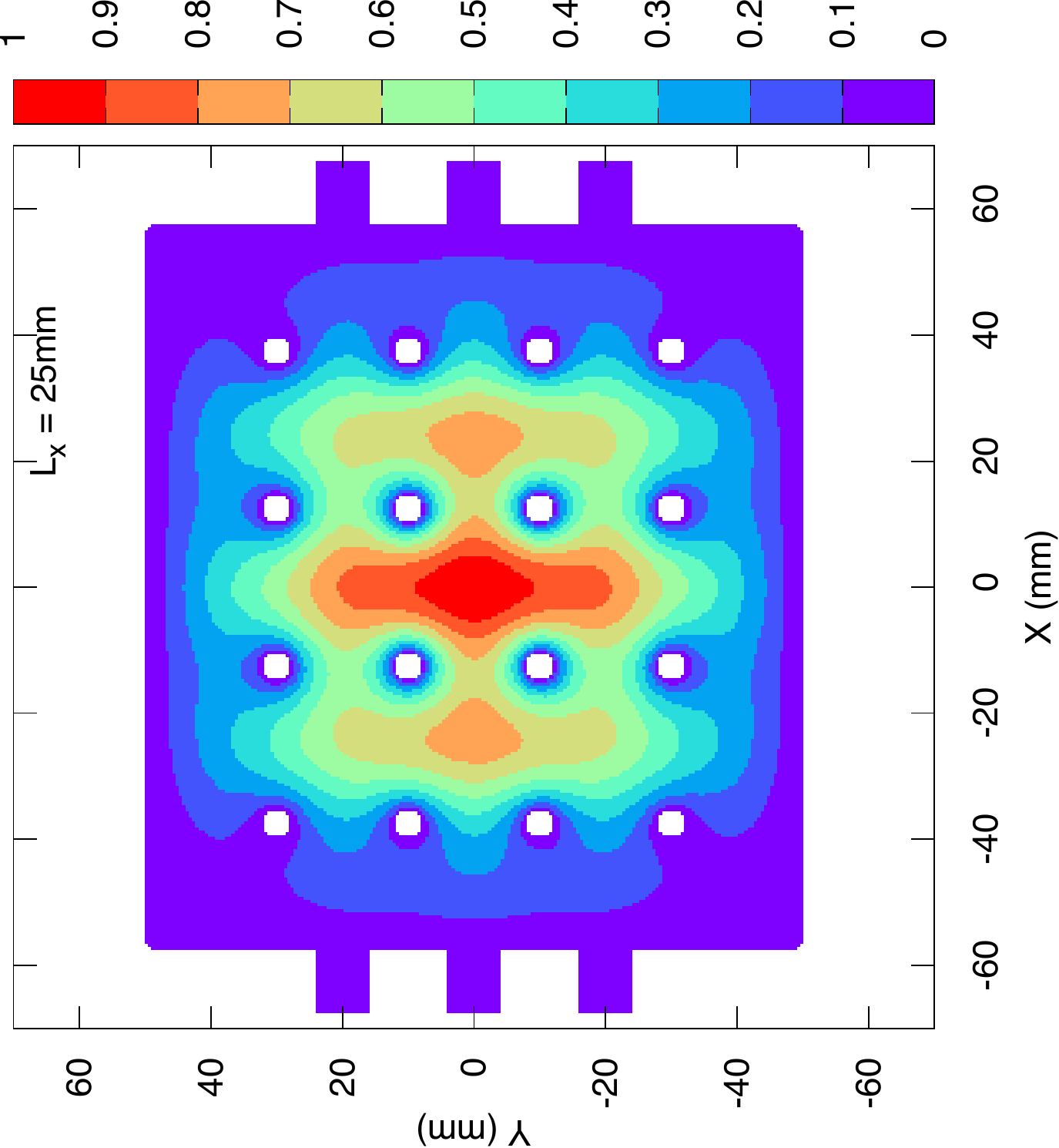}} \label{fig:Lx25mm}
\caption{Calculated magnitude of electric field in the lowest resonance mode on $xy$ plane, $|{\bm E}(x, y, z=0)|$.  Normalization of the field strength is set to 1 at the cavity center, $(x, y, z)=(0, 0, 0)$.
 From (a) to (l), $L_x$ is changed from 14 mm to 25 mm in 1 mm step. There is no node in the electric field shape. 
 It is similar to ${\rm TM_{010}}$.}
\label{fig:Contour}
\end{center}
\end{figure}

Two cavities optimized for about 5 to 7 GHz were developed in this study.
One was a smaller size (Cavity A), but the resonant frequency can be tuned. 
Another was a bigger (Cavity B) to demonstrate the feasibility of a large volume cavity,
but it was not equipped with a frequency tuning mechanism.
In this frequency region, the period of photonic crystals is around 15 - 25 mm.
The base design of Cavity A is a rectangular shape with a length and a width of 100 mm 
where 16 cylinders with a diameter of 4 mm are arranged in a grid pattern as shown in Fig. \ref{fig:designAndVector}(a).
The distances between the  grids in the $y$ direction, $L_y$,
and the ones between the cavity inner walls and the grids at both ends are set to the constant, 20 mm.
On the other hand, the grid distance in $x$ direction, $L_x$, is changed to tune the resonance frequency. 
The thickness of the cavity in the $z$ direction is 10 mm. The right panel of Fig. \ref{fig:designAndVector} 
shows the simulated electric field vector on the $xy$ plane, ${\bm E}(x, y, z=0)$, in the lowest resonance mode at $L_x$ = 20 mm.
The calculations were conducted by the finite element method, 
the eigenmode solver of Ansys HFSS.
Contour plots of the magnitude of the electric field strength, $|{\bm E}(x, y, z=0)|$, when $L_x$ is changed from 15 to 25 mm are shown in Fig. \ref{fig:Contour}.
As shown in the figures, no-node mode, like ${\rm TM_{010}}$ mode, can be obtained by the photonic crystal structure.

\begin{figure}[!h]
\begin{center}
\centering{\includegraphics[width=2.5in, angle=0, clip]{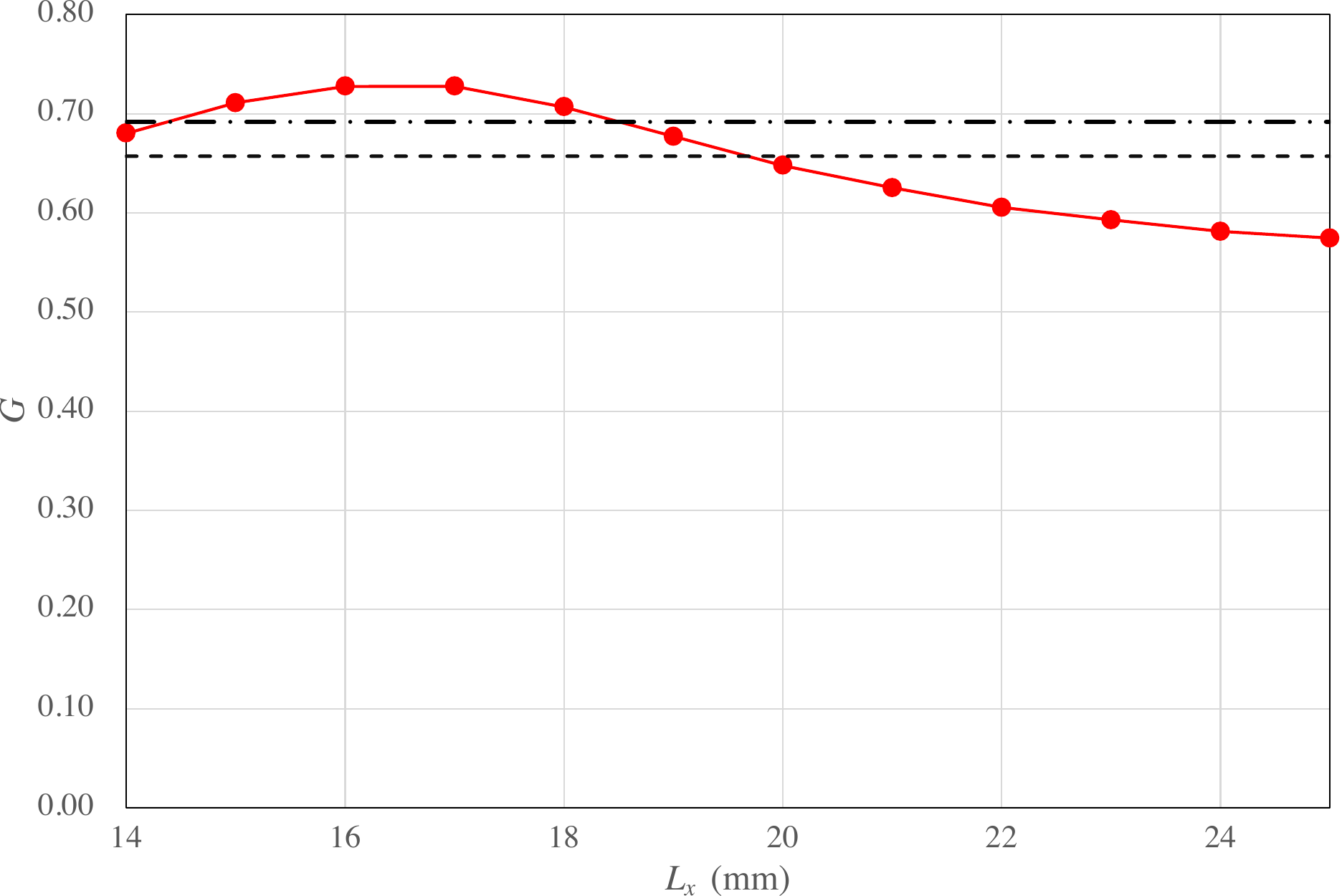}} 
\caption{Calculated $G$-factor vs $L_x$. The $G$-factor at each case is calculated 
from the results of the electric field 
by the eigenmode solver in the finite element method (ANSYS HFSS). 
The simulated results are shown by red circles. The $G$-factors of  ${\rm TM_{010}}$ mode in the cylindrical- and the rectangular cavity, that is, $G=0.69$ and 0.65,
are also drawn by a dash-dot line and a broken line.}
\label{fig:gFactor}
\end{center}
\end{figure}

We also calculate the $G$-factor under the assumption that the uniform magnetic field is applied in the $z$ direction.
The results in Cavity A are plotted in Fig. \ref{fig:gFactor}, 
and the one for Cavity B is $G=0.63$.
The $G$-factor of Cavity A
at $L_x=20$ mm is very close to the ideal rectangular cavity's, $G=0.65$,
but it becomes slightly smaller for longer $L_x$.
For shorter $L_x$, between 15 -- 18 mm, 
it increases and is more than that of the cylindrical cavity,
$G=0.69$.
Looking at (c) and (d) of Fig. \ref{fig:Contour}, the electric field becomes large near the wall, $|x|\sim30$ mm.
It is inferred that the distance from the grid to the wall, 20 mm,  is particularly suitable for these resonance frequencies,
and is considered that the $G$-factor is high due to the electric field distribution.
Thus, the lowest mode of the DRiPC cavity is found to have good features of being tunable by $L_x$ and having no node and high $G$-factor.

We also consider the effects caused by imperfection of the grid positions,
which can cause mode localization in the cavity and lead degradation of the $G$-factor \cite{bib:modeLocalization}.
To see the effect, we conducted simulations where one of the poles is mis-aligned.
The shifted pole position is $(x, y)=(+1 {\rm mm}, -1 {\rm mm})$ and the displacement is $\Delta y=+0.3{\rm mm}$ and $+0.5 {\rm mm}$ in $y$ direction.
No significant mode localization is found even at $\Delta y=+0.5 {\rm mm}$.
According to a manufacturer, the tolerance of the grid positions is less than 0.2 mm.
Combining these information, we could expect no strong mode localization. 

To realize the DRiPC cavity illustrated in Fig. \ref{fig:designAndVector}  (a), 
we fabricated a cavity made of oxygen-free copper (JIS 1020) consisting of the following three components.
Component A was an end part of the cavity with three holes (8.4 mm in diameter) parallel to $x$ axis for equipping microwave antennas, etc.
Component B was a 100 mm $\times$ 10 mm hollow hole with 10 mm thickness,
in which four cylinders with a diameter of 4 mm are arranged in y- direction.
With these four parts, 16 poles were arranged in a rectangular grid pattern.
The last one, Component C, was a set of plates with a rectangular hole (100 mm $\times$ 10 mm)
and inserted between Component B to adjust $L_x$. 
Three 10 mm-thickness and 30 1 mm-thickness were prepared as Component C.
$L_x$ can be changed by combining 10 mm and 1 mm plates.
In the actual measurements, 
only 1 mm spacers were used for $L_x \le 20$ mm ("Short setup").
At $L_x \ge 20$ mm, both of 1 mm and 10 mm thick  were used ("Long setup").
Table \ref{tbl:cavity} shows the actual configuration of Component C inserted between the $x$-grid and actual $L_x$ measured with a caliper.
\begin{table}[!h]
\caption{Cavity arrangements. Setup names are for future convenience. "S" and "L" mean "Short" and "Long", respectively. For $L_x \ge 20$ mm (Long setup), we use the 10 mm plates with 1 mm plates. For $L_x\le 20$ mm (Short setup), only 1 mm plates are used.}
\label{tbl:cavity}
\centering
\begin{tabular}{|c|c|c|c|}
\hline
Setup name & \multicolumn{2}{|c|}{Used Component C} & Measured \\
\cline{2-3} 
  &10 mm plate & 1 mm plates &  $L_x$ (mm) \\
\hline
S-14&	 & 4 & 13.9 \\
S-15&	 & 5 & 14.9 \\
S-16&	 & 6 & 15.9 \\
S-17&	0 & 7 & 16.9 \\
S-18&	 & 8 & 17.9\\
S-19&	 & 9 & 19.1 \\
S-20&	 & 10 & 20.0 \\
\hline
L-20 & & 0 & 19.8 \\
L-21&  & 1 & 20.9 \\
L-22 &1 & 2 & 21.8 \\
L-23&   & 3 & 22.9 \\
L-24&  & 4 & 23.9 \\	
L-25& & 5 & 25.0 \\
\hline
\end{tabular}
\end{table}

\begin{figure}[!h]
\begin{center}
\centering
{\includegraphics[width=2.4in, angle=0, clip]{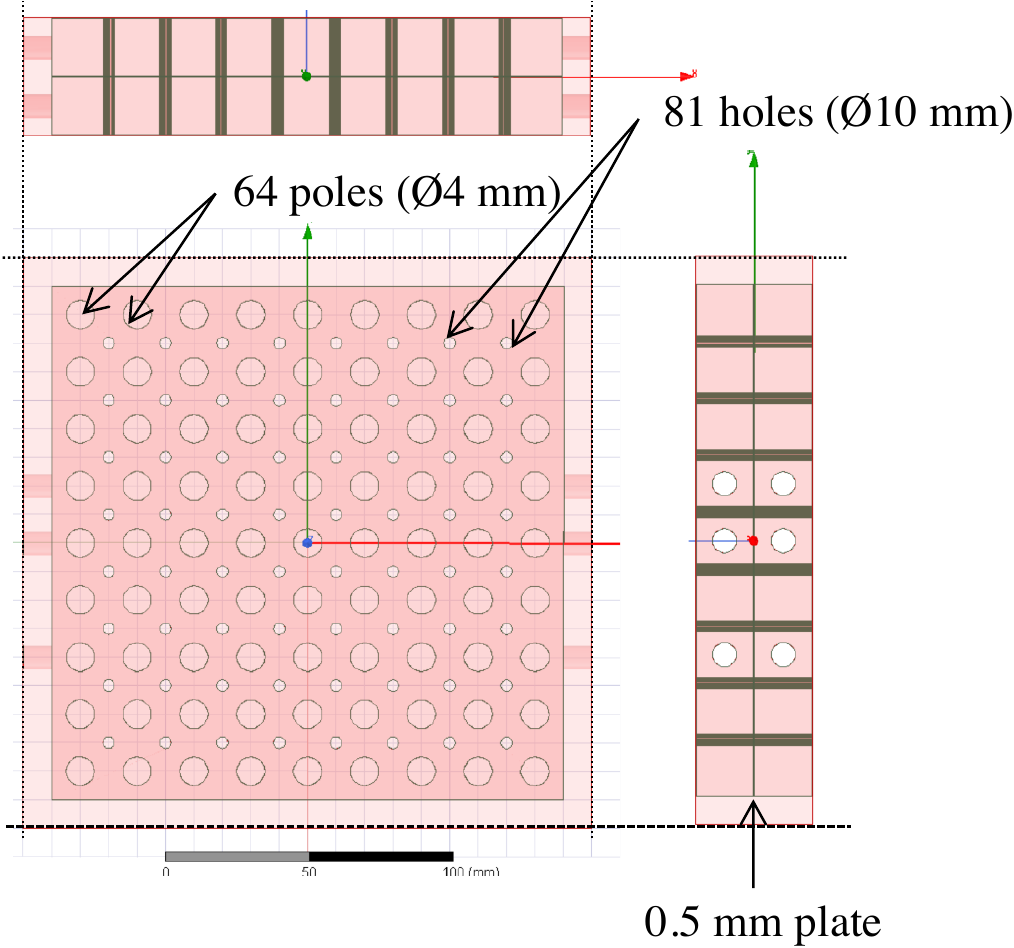}}
%
{\includegraphics[width=2.2in, angle=0, clip]{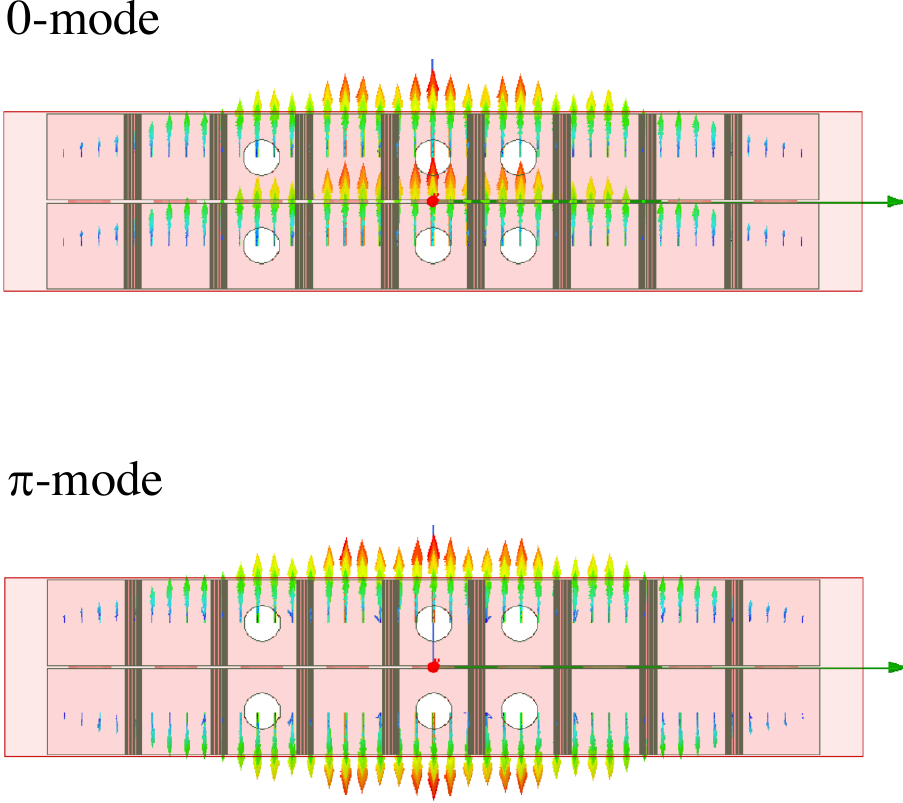}}

\caption{A three view drawing of the larger DRiPC cavity (Left). The dimensions of this cavity, Cavity A, are enlarged: 180 mm-length and  width and 20 mm-thickness. 
These two were stacked and separated by a 0.5 mm-thick plate, 
where there are 81 10 mm-diameter holes for microwave coupling. 
The photonic crystal structure was constructed by 64 4 mm-diameter poles.
All were made of oxygen-free copper.
The right top and bottom panel shows 
the simulated electric field on the plane at $z=\pm10.25$ mm in the lowest (0-mode) and the 2nd lowest resonance mode ($\pi$-mode), respectively.
}
\label{fig:3ViewLCavity}
\end{center}
\end{figure}

We also prepared a larger cavity made of oxygen-free copper to show the feasibility of the DRiPC cavity to extend to a larger size.
We expanded the size in the $x$ and $y$ direction to 180 mm 
and the $z$ direction to 20 mm
and piled up in two stages.
The two cavities are separated 
by 0.5 mm thick oxygen-free copper which had 81 10 mm-diameter holes for microwave coupling between the upper and the lower side. 
We arranged 64 oxygen-free copper cylinders with a diameter of 4 mm at $L_x = L_y = 20$ mm 
in a rectangular shape. 
In this cavity, the volume is reached to $1.3 \times 10^3$  ${\rm cm^3}$, 
but the frequency tunability is omitted.
The cavity structure is illustrated in Fig. \ref{fig:3ViewLCavity}.
The reason to use the coupled cavity is to avoid the possibility of complicated mode structure due to higher modes with nodes in $z$-direction. 
In this coupled cavity, the lowest frequency mode is the one in which the lowest ${\rm TM_{010}}$-like mode in each cavity resonates in the same phase (0-mode), and the second lowest is the mode it oscillates in opposite phase ($\pi$ mode), as illustrated in the right panel of Fig. \ref{fig:3ViewLCavity}.
As the $G$-factor vanishes in $\pi$-mode,
0-mode will be used in the axion searches.

\section{Measurements and results}\label{sec:measAndResults}
\subsection{Resonance frequency and $Q$-value measurement in the 100 mm $\times$ 100 mm $\times$ 10mm-cavity}\label{subsec:resonalceFreq}
\begin{figure}[!h]
\begin{center}
\centering
{\includegraphics[width=4.5in, angle=0, clip]{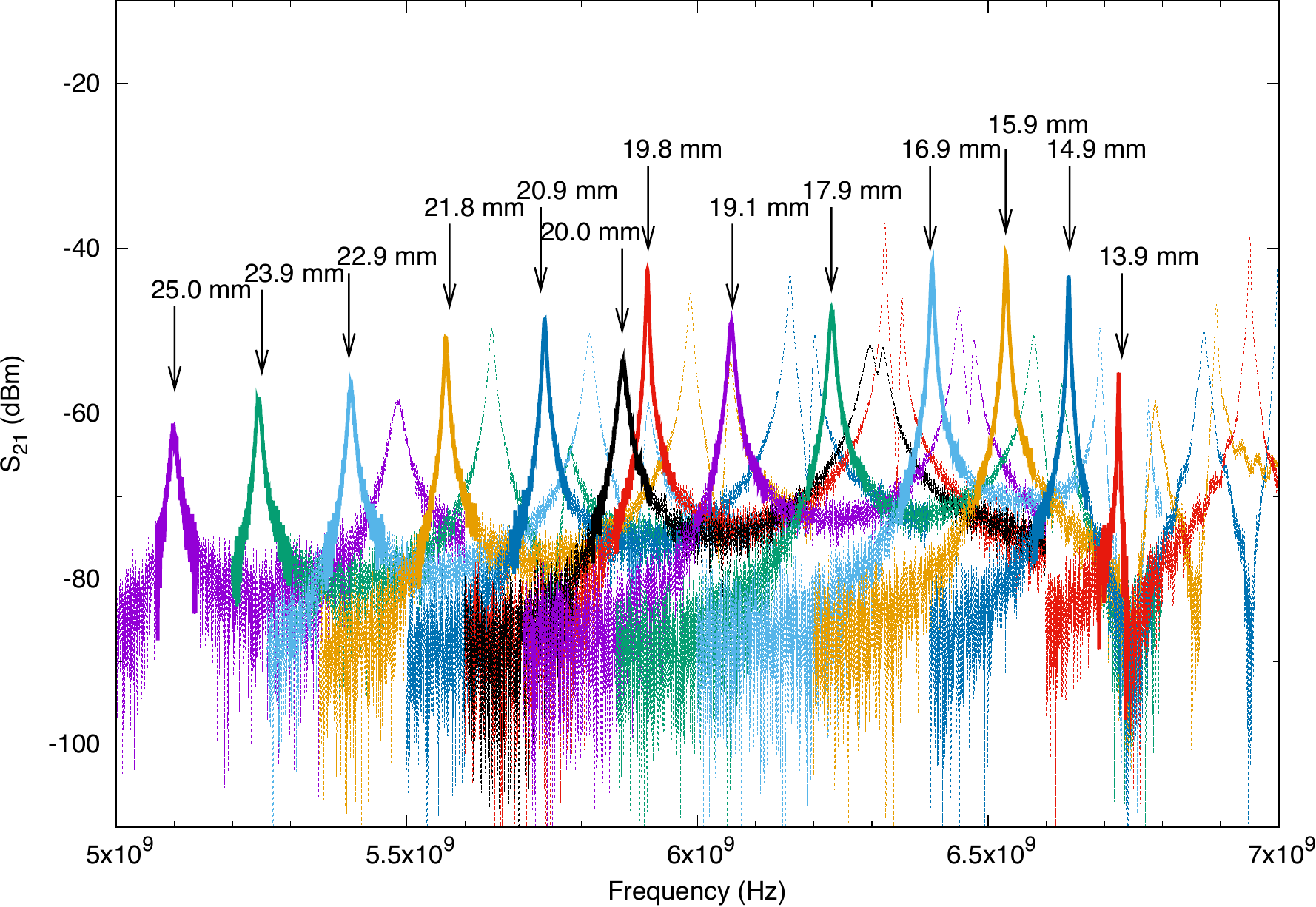}} 
\caption{The spectrum of $S_{21}$ at each $L_x$. $L_x$ is changed from around $\sim14$ mm to 25 mm by 1 mm step. The peak position of ${\rm TM_{010}}$-like mode at each $L_x$ value is pointed by an arrow with a legend. }
\label{fig:freqLx}
\end{center}
\end{figure}

We firstly investigated the relationship between $L_x$ and the resonance frequencies of Cavity A, 100 mm $\times$ 100 mm $\times$ 10mm-cavity.
Two loop antennas were set at the holes at $y=10$ mm in Component A. The direction of the loop was faced to $x$-direction to be sensitive to ${\rm TM_{010}}$ mode.
The three lowest resonance-frequencies in the cavity were measured by the transmission spectrum, $S_{21}$, 
using a vector network analyzer (VNA), E5063A manufactured by Keysight Technologies.
The system was calibrated by the standard 2-port calibration method.
The $S_{21}$ spectrum corresponding to each $L_x$ is shown in Fig. \ref{fig:freqLx}.
According to the simulation results, 
the peak with the lowest frequency is the fundamental ${\rm TM_{010}}$-like mode.
The second and third ones are higher modes that have one node.
An important feature is the absence of peaks in the low frequency domain, 
which leads that mode crossing never happens in principle.

\begin{figure}[!h]
\begin{center}
\centering
{\includegraphics[width=4.5in, angle=0, clip]{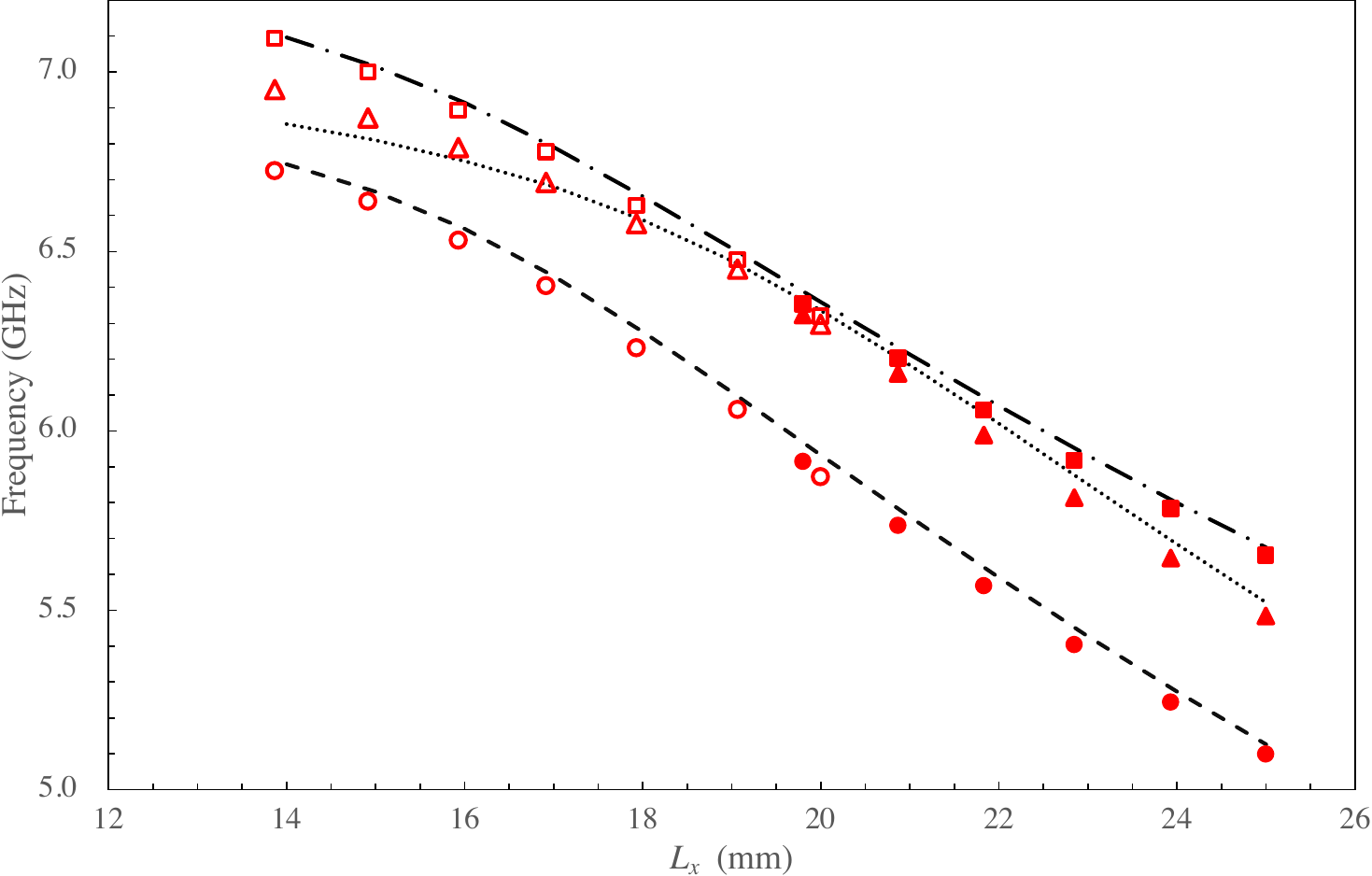}}
\caption{The resonance frequencies vs $L_x$ in Cavity A.
The lowest frequency is the fundamental ${\rm TM_{010}}$-like mode. The measured and the simulated data are indicated by circles and a broken line. The following two are high-order modes.
These two modes have one node in the $xy$ plane.
Measured frequencies are plotted by triangular and square marks.
Simulated ones are drawn by a dotted line and a dotted-broken line. 
The closed marks and the open ones in the plot are used to distinguish "Long" and  "Short" setup, respectively.}
\label{fig:resonanceF}
\end{center}
\end{figure}

Figure \ref{fig:resonanceF} represents the resonance frequencies as a function of  $L_x$.
The points are the measurement data, and the lines are the results of the simulation.
The closed marks represent the data of "Short" setup in Table \ref{tbl:cavity},
and the open marks "Long" setup, respectively. 

Although there are certain deviations between the absolute values of the two data sets, the data and the simulation results in all three modes show excellent agreements on the frequency dependence on $L_x$.
The only exception is the 2nd mode at $L_x \sim14$ and 15 mm.
As for the lowest-order ${\rm TM_{010}}$-like mode, 
which is used in axion searches and therefore of most interest, 
the difference between the data and the simulation was about 60 MHz at $L_x = 20$ mm.
It could be explained by some factors which could not be taken into account in the simulations, 
for example, the shape of the welded part of the grids, and/or the machining accuracy, and so on.
The difference is not so important because the resonance frequency is scanned to match unknown axion mass in actual searches.

Then, we focus on the lowest resonance frequency.
As lengthening $L_x$ from the base design ($L_x=20$ mm) to $L_x=25$ mm,
it was almost linearly tuned from 5.87 GHz to 5.10 GHz, that is, $-13.2$\%.
In the direction of shortening $L_x$ from 20 mm to 13.9 mm,
on the other,
the frequency was reached to 6.72 GHz, $+14.5$\%, 
although loosing linear dependency when $L_x$ is less than $\sim15$ mm.
In total 27.7\% frequency range was covered.
Since a wide frequency range is searched in the axion hunting,
this broad coverage is a remarkable feature.

In the range of $L_x=17 \sim 25$ mm, 
it showed a proportional behavior, 
and changing rate per unit length is $\Delta f/\Delta x={\rm 164}$ MHz/mm.
The counter partner in the simulation is ${\rm 154 }$ MHz/mm.
It is also a very good agreement.
The frequency step at the axion search is 
$\Delta f \sim f/Q_c$, where $Q_c$ is loaded $Q$-value of the cavity. 
If we set $\Delta f/f \sim 10^ {-6}$, corresponding to the axion energy width,
the signal is expected to be largest and most efficient. 
Specifically, in our cavity, it is about 5 to 6 kHz.
With the obtained value, ${\rm 164 }$ MHz/mm, 
it found difficult to fine-tune the frequency by adjusting $L_x$:
For example, the minimum accuracy of  a commercially available micrometer, 
or a positioning device, is $\Delta x \sim 0.5\, \mu$m 
and the resultant minimum frequency-change is estimated to  $\Delta f \sim 81$ kHz. 
It means that another fine frequency-tuning mechanism is required.
This fine tuning mechanism also needs to cover continuously the frequency gap between discrete frequencies in the current design, that is, inserting plates.
One of the ideas is to utilize a bead. 
As we seen in Sec. \ref{subsec:eField}, 
the resonant frequency gets lower when a metal bead is introduced.
When a dielectric bead, it gets higher.
One candidate of the dielectric material is high-purity sapphire, 
whose microwave properties are well studied and the microwave loss is small enough. 
The actual implementation of such a fine tuning mechanism is left for future studies,
but,
as can be seen from Fig. \ref{fig:resonanceF}, 
the frequency can be tuned in a very wide range effectively by adjusting the grid spacing, $L_x$.

Loaded $Q$-value at each setup was also obtained from the $S_{21}$ spectrum,
$Q_L = f/( f'-f'' )$, 
where $f$ is the resonance frequency 
and $f'$ and $f''$ are  the frequencies where $S_{21}$ 
is lower than -3.0 dB from the peak.
Unloaded $Q$-values are, then, calculated by $Q_0 = Q_L\times  \frac{1}{1-10^{S_{21}  {(\rm dB)/20}}}$ \cite{bib:q0Calc}.
In this experimental setup, we adjusted to make the coupling to the cavity very weak, and the correction from $Q_L$ to $Q_0$ is very small.
The results of $Q_0$ are tabulated in Table \ref{tbl:qValue}.
\begin{table}[!h]
\caption{Measured $Q_0$ for each setups in the 100 mm $\times$ 100 mm $\times$ 10 mm-cavity.}
\label{tbl:qValue}
\centering
\begin{tabular}{|c|c|c|c|}
\hline
Setup name & $L_x$ (mm) & Inserted number of plates & $Q_0$ \\
\hline
S-14 & 13.9 & 4 & $2.8\times 10^3$ \\
S-15 & 14.9 & 5 & $2.1\times 10^3$ \\
S-16 & 15.9 & 6 & $1.6\times 10^3$ \\
S-17 & 16.9 & 7 & $1.3\times 10^3$  \\
S-18 & 17.9 & 8 & $7.4\times 10^2$  \\
S-19 & 19.1& 9 & $6.9\times 10^2$  \\
S-20 & 20.0 & 10 & $4.8\times 10^2$  \\
\hline
L-20 & 19.8 & 1 & $1.4\times 10^3$  \\
L-21 & 20.9 & 2 & $9.6\times 10^2$  \\
L-22 & 21.8 & 3 & $9.3\times 10^2$  \\
L-23 & 22.9 & 4 & $6.0\times 10^2$  \\
L-24 & 23.9 & 5 & $6.0\times 10^2$  \\
L-25 & 25.0 & 6 & $4.7\times 10^2$  \\
\hline
\end{tabular}
\end{table}
As expected, $Q_0$ decreases as the number of inserted plates increases.
Even if the number of plates is small, such as the setup "L-20", 
only moderate $Q_0$ was obtained.
Since this cavity was made of oxygen-free copper (JIS C1020), 
it is possible to use higher-purity copper to increase the $Q$-value.
Some devices to ensure much more solid electric-contact between the plates could also be adopted.
It is also  considered to introduce technologies described in refs. \cite{bib:scCavityNbTi, bib:scCavityYBCO}. 
But the improvement of the $Q$-value is future work.

\subsection{Measurement of the electric field profile in the 100 mm $\times$ 100 mm $\times$ 10 mm-cavity }\label{subsec:eField}
\begin{figure}[!h]
\begin{center}
\centering
{\includegraphics[width=4.5in, angle=0, clip]{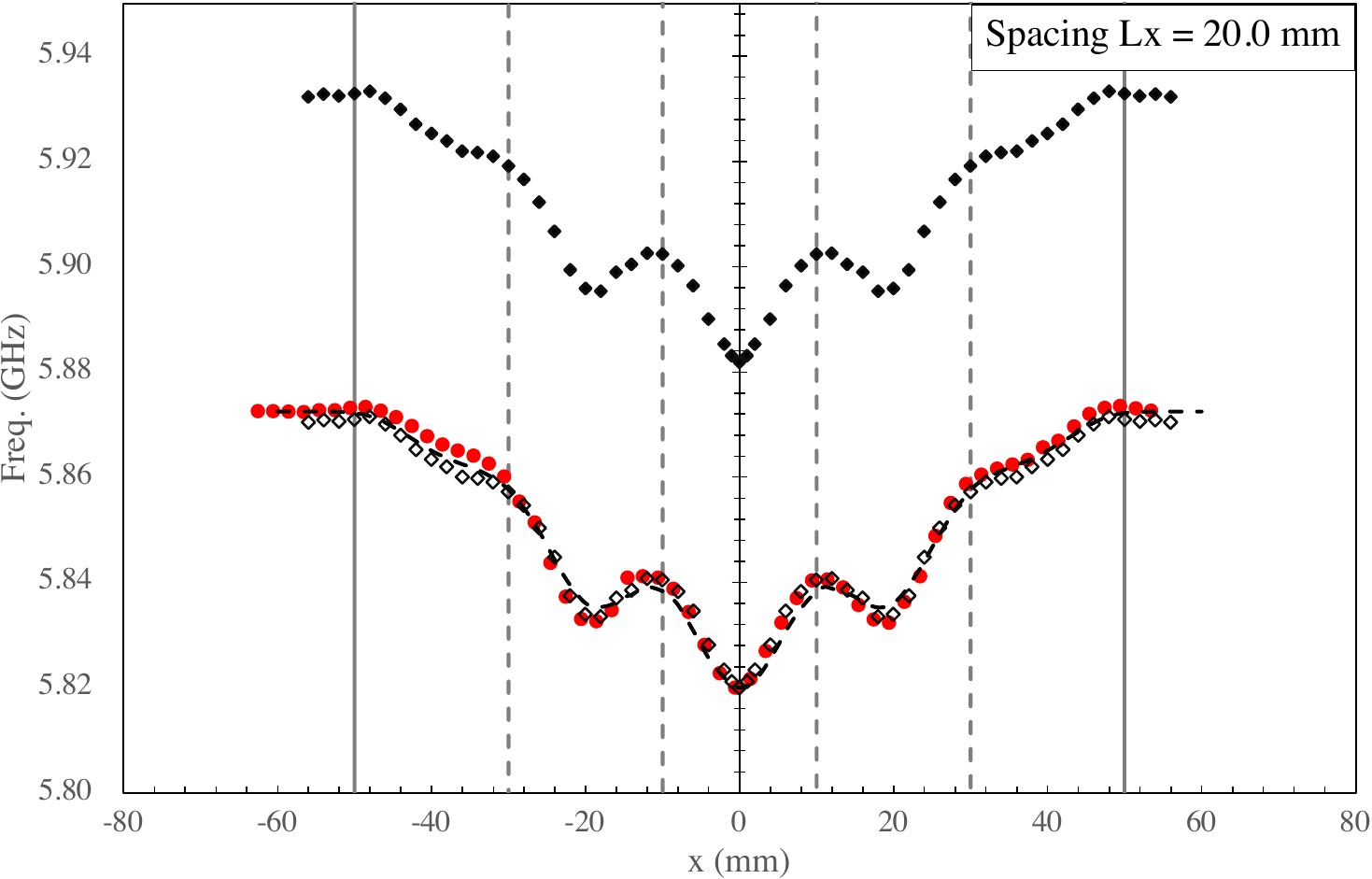}}
\caption{The resonance frequency changes by the introduction of the copper bead. 
The experimental results and the simulated ones are plotted by red circles and black diamonds, respectively. 
There is about a 60 MHz gap between the two.
The simulation results are shifted to match the experimental value at $x=0$ mm, 
and are plotted by open diamond. 
There is an excellent agreement between them.
We also estimated the frequency shift by $E(x)^2$ in Eqn. \ref{eqn:freqShift}. 
In the estimation, the coefficient $a$ 
was determined 
so that $\Delta f$ at $x=0$ is reproduced the experimental value. 
With that $a$, we estimated $\Delta f$ at  $x\ne0$ points. 
The estimated value is drawn by a black dash line.
The two thick-gray vertical line indicate the position of the inner cavity wall.
The four thin-gray broken lines correspond to the positions of the grids.
We can see that the electric field is lower around the poles arranged in the lattice shape.
}
\label{fig:fieldX20wS}
\end{center}
\end{figure}

The electric field profile in Cavity A at each $L_x$ 
was measured by the bead pull method.
In the bead pull method,
the strength of an electric field is measured
by introducing a small dielectric or conducting bead into a cavity 
and monitoring the frequency change due to its perturbation effect \cite{bib:cernAS}.
When a non-magnetic conductor is introduced in ${\rm TM_{010}}$ mode,
in particular,
the frequency shift and the electric field strength are related by the following equation:
\begin{equation}\label{eqn:freqShift}
	\frac{\Delta f}{f}= \frac{\Delta U(x)}{U}  \simeq  a \times E(x)^2, 
\end{equation}	
where $\Delta U(x)$, $U$ and $a$ are the electric field energy at position $x$, the total stored energy in the cavity and a constant, respectively.
$E(x)$ is the electric field at the introduced perturbation position $x$.
In the experiment, a copper ball with 5.5 mm diameter was introduced along the $y$ axis using a nylon fishing line,
and the frequency change was measured at 2 mm intervals.
Figure \ref{fig:fieldX20wS} shows the experimental results at $L_x = 20$ mm.
In Fig. \ref{fig:fieldX20wS}, 
the simulation results of the resonance frequency when the bead is introduced at each $x$ are also potted by closed diamonds.
Since there is about a 60 MHz difference in the measured frequency and simulated one,
the latter is shifted so that both of the resonance frequencies at $x = 0$ mm matches, 
and it is depicted in open diamonds.
The measured and the simulated results agree with each other with very high accuracy, 
indicating that the electric field distribution of the DRiPC cavity is ${\rm TM_{010}}$-like,
and therefore the cavity is suitable for halo-scope experiments.
In addition, frequency shift by a simple calculation in the last part of Eqn. \ref{eqn:freqShift} is also drawn by a black broken line in the Fig. \ref{fig:fieldX20wS}.
Here, the constant $a$ is scaled so that the resonance frequency before the bead introduction and the one at $x = 0$ mm match with the measured values.
It can be seen that this calculation well reproduces the simulation result with the small bead.
In the following, therefore, we will discuss by comparing this simple calculation result with the experimental data.

\begin{figure}[!h]
\begin{center}
\centering
\subfloat[] [$L_x$=15.9 mm]{\includegraphics[width=1.98in, angle=0, clip]{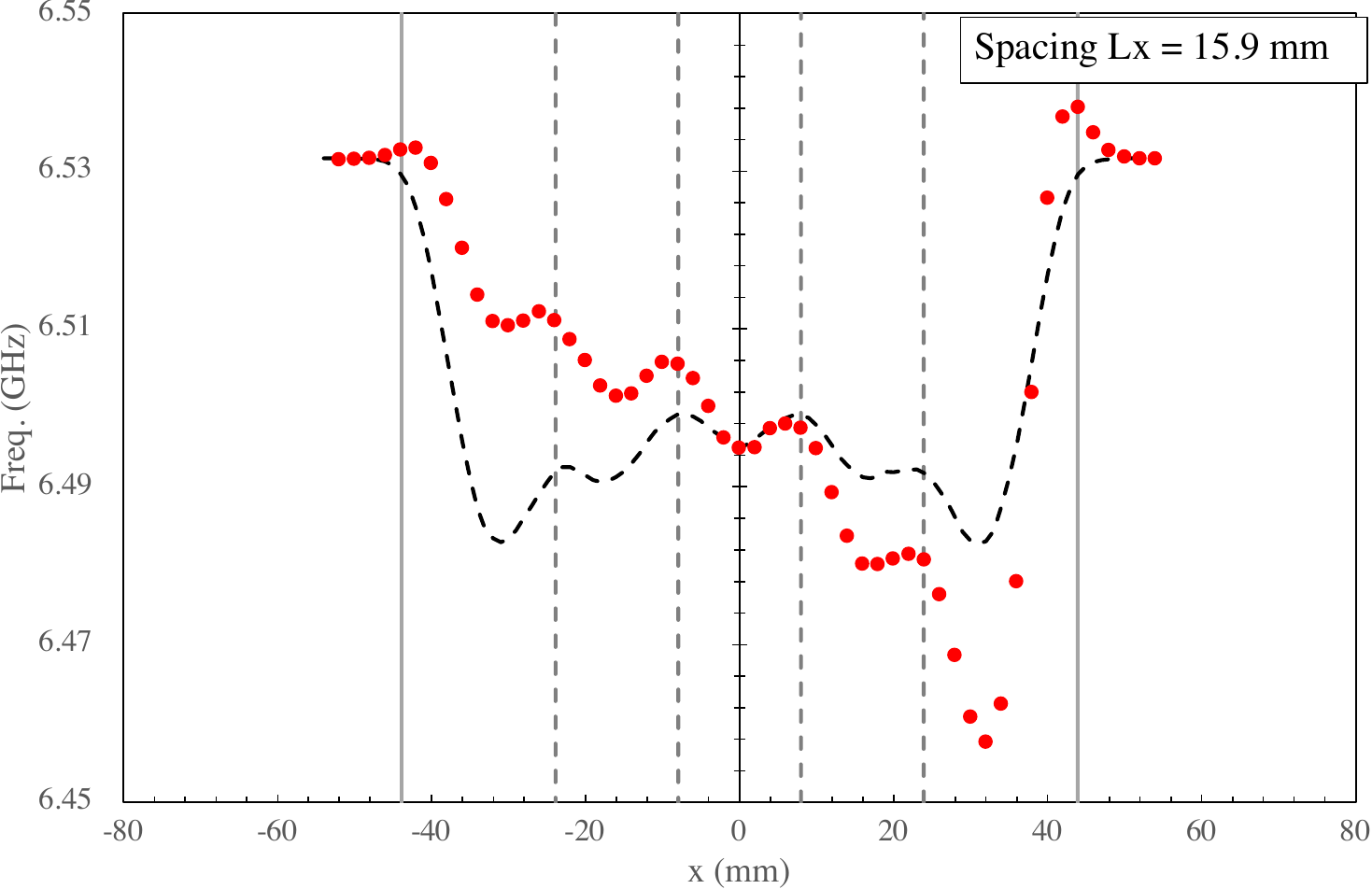}} \label{subfig:fLx16mm}
\subfloat[] [$L_x$=16.9 mm]{\includegraphics[width=1.98in, angle=0, clip]{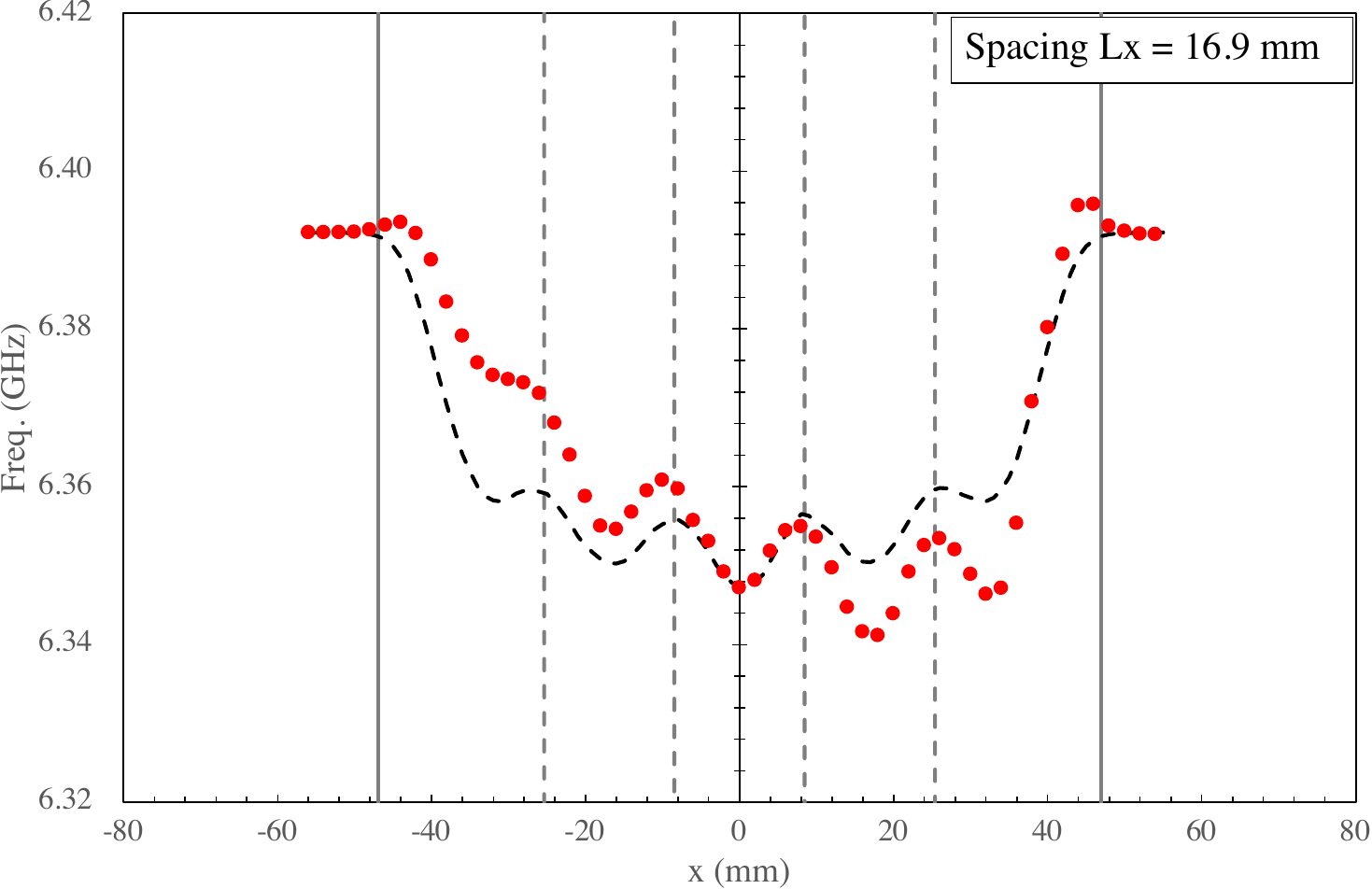}} \label{subfig:fLx17mm}
\subfloat[] [$L_x$=17.9 mm]{\includegraphics[width=1.98in, angle=0, clip]{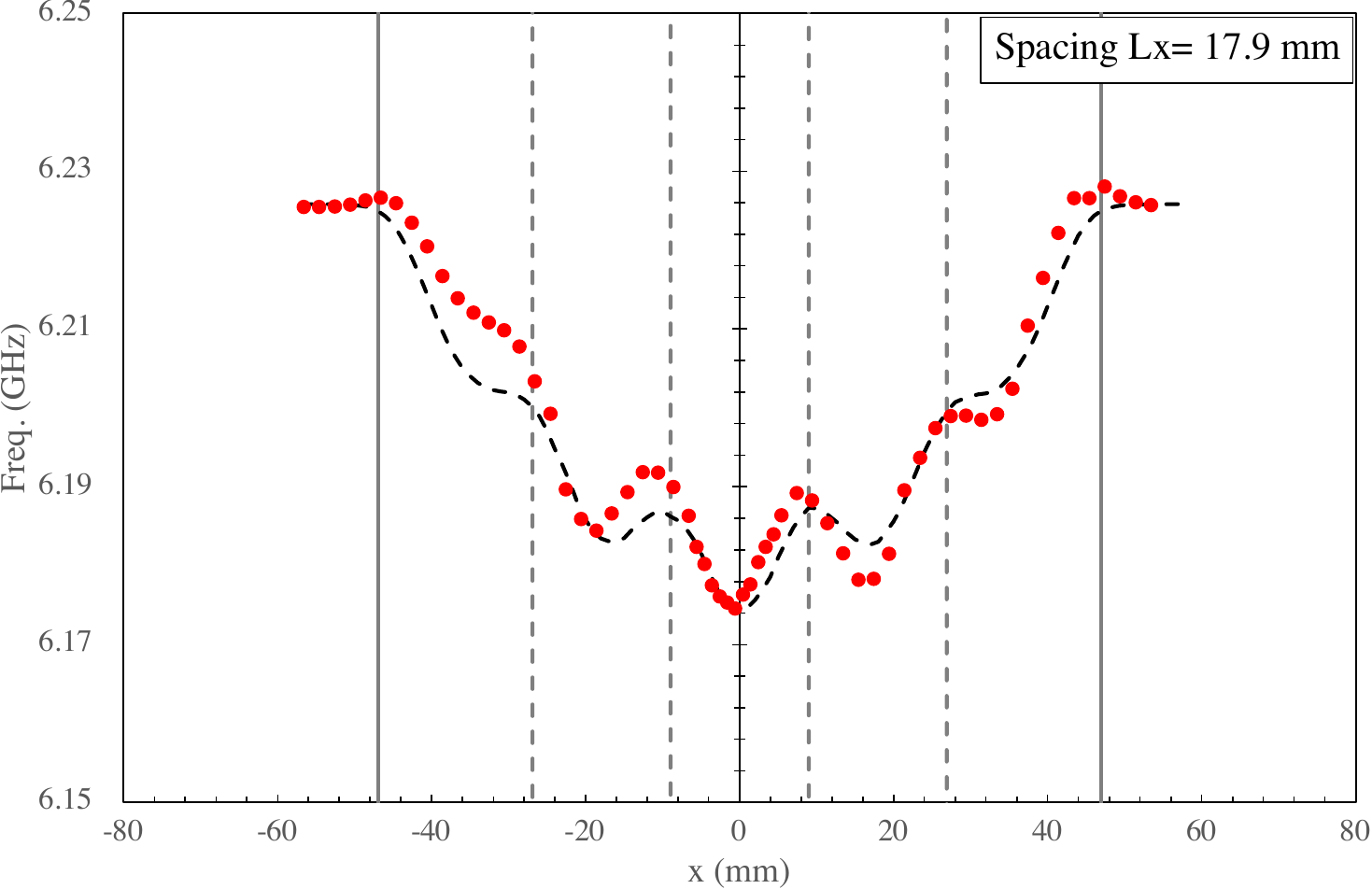}} \label{subfig:fLx18mm}

\subfloat[] [$L_x$=19.1 mm]{\includegraphics[width=1.9in, angle=0, clip]{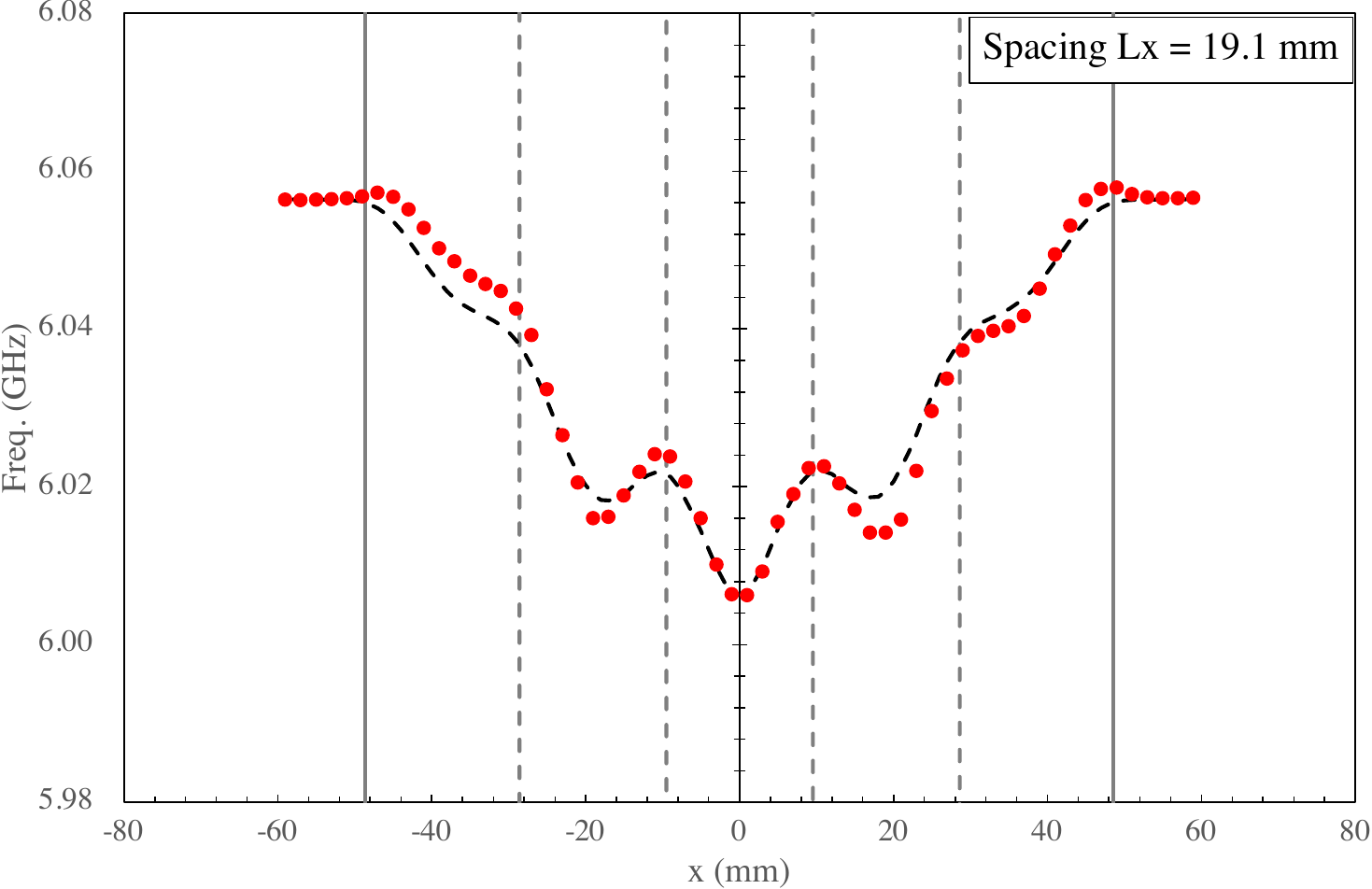}} \label{subfig:fLx19mm}
\subfloat[] [$L_x$=20 mm]{\includegraphics[width=1.98in, angle=0, clip]{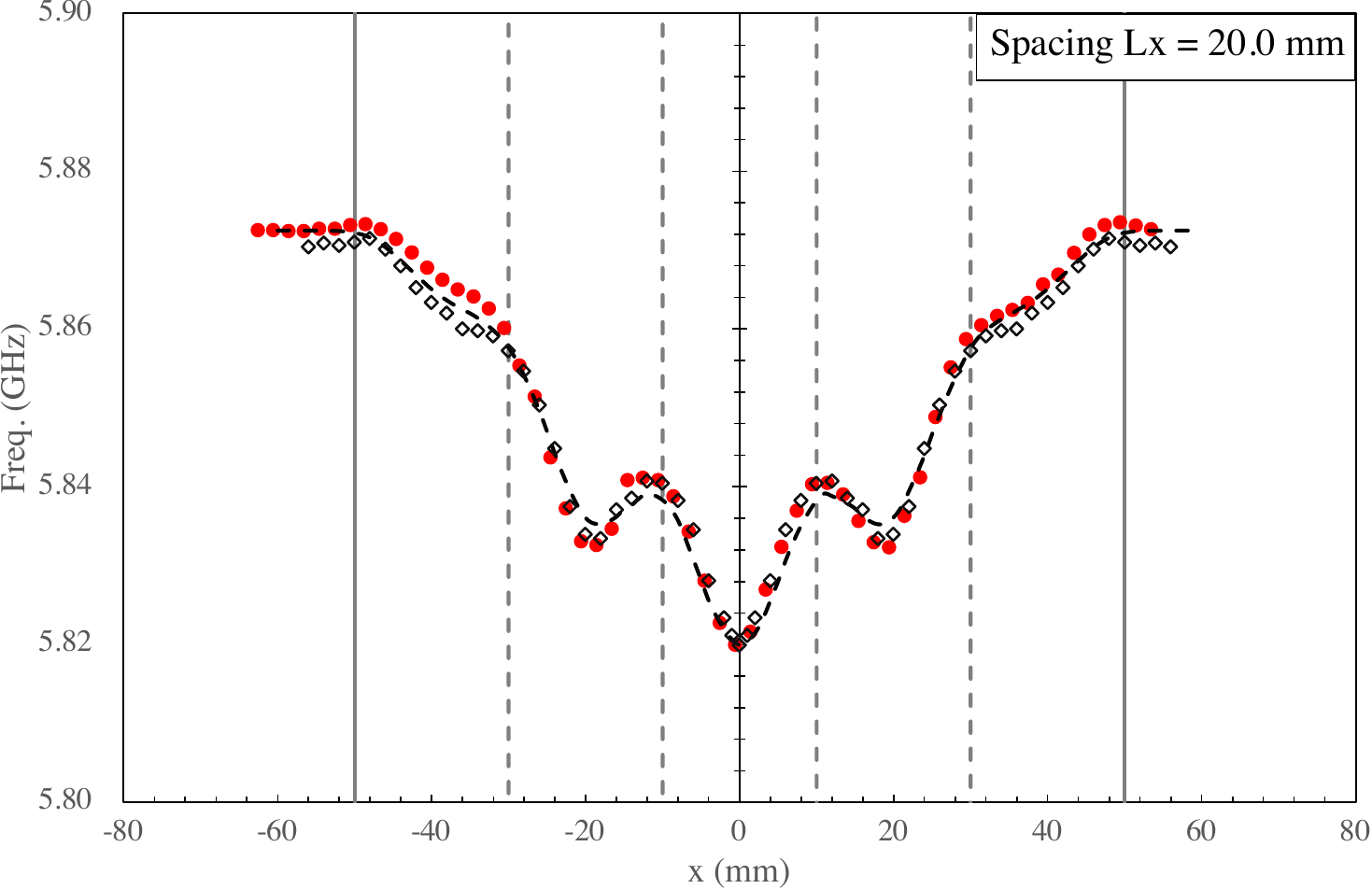}} \label{subfig:fLx20mm}
\subfloat[] [$L_x$=19.8mm]{\includegraphics[width=1.98in, angle=0, clip]{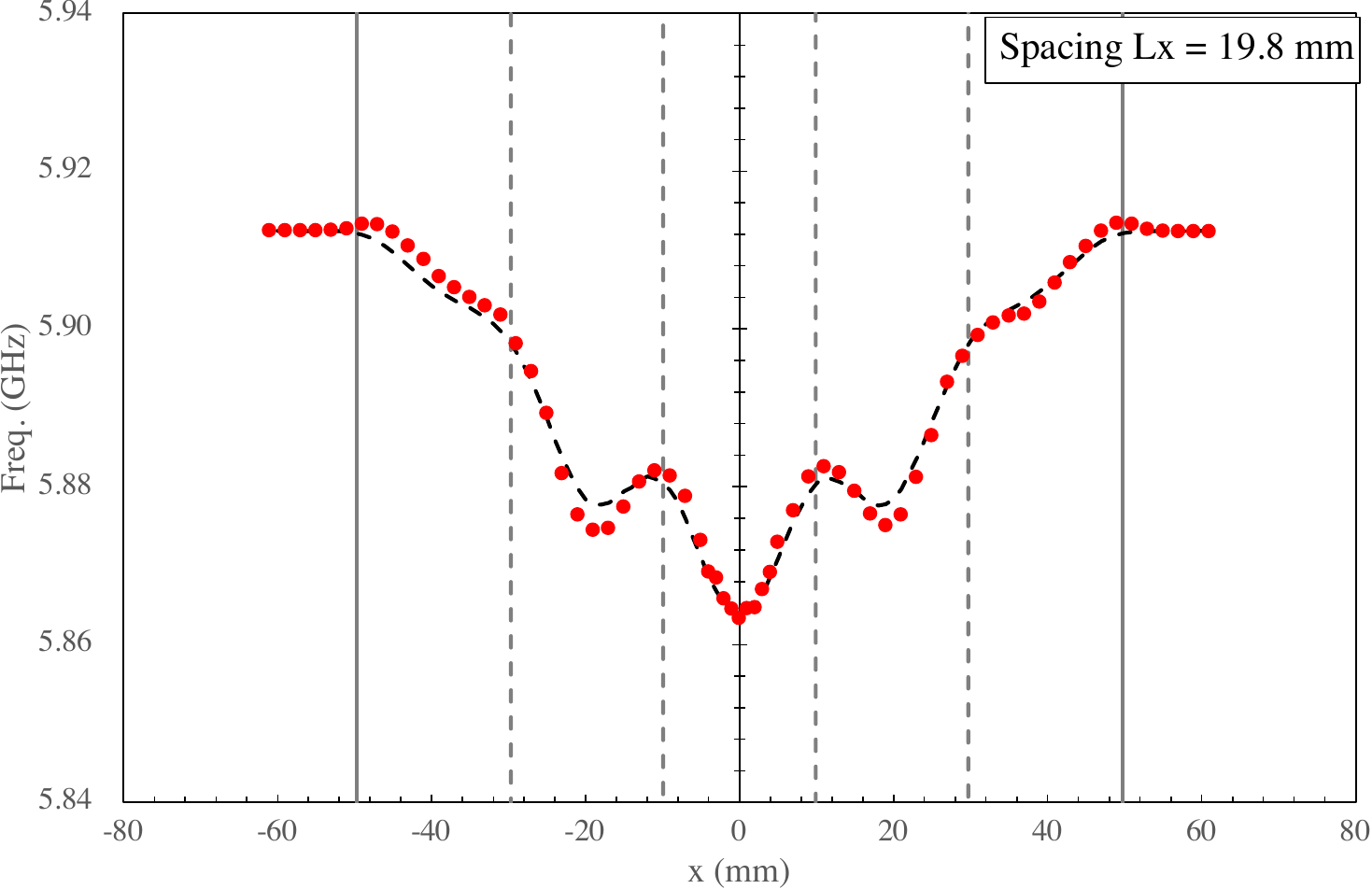}} \label{subfig:fLx19p8mm}

\subfloat[] [$L_x$=20.9 mm]{\includegraphics[width=1.98in, angle=0, clip]{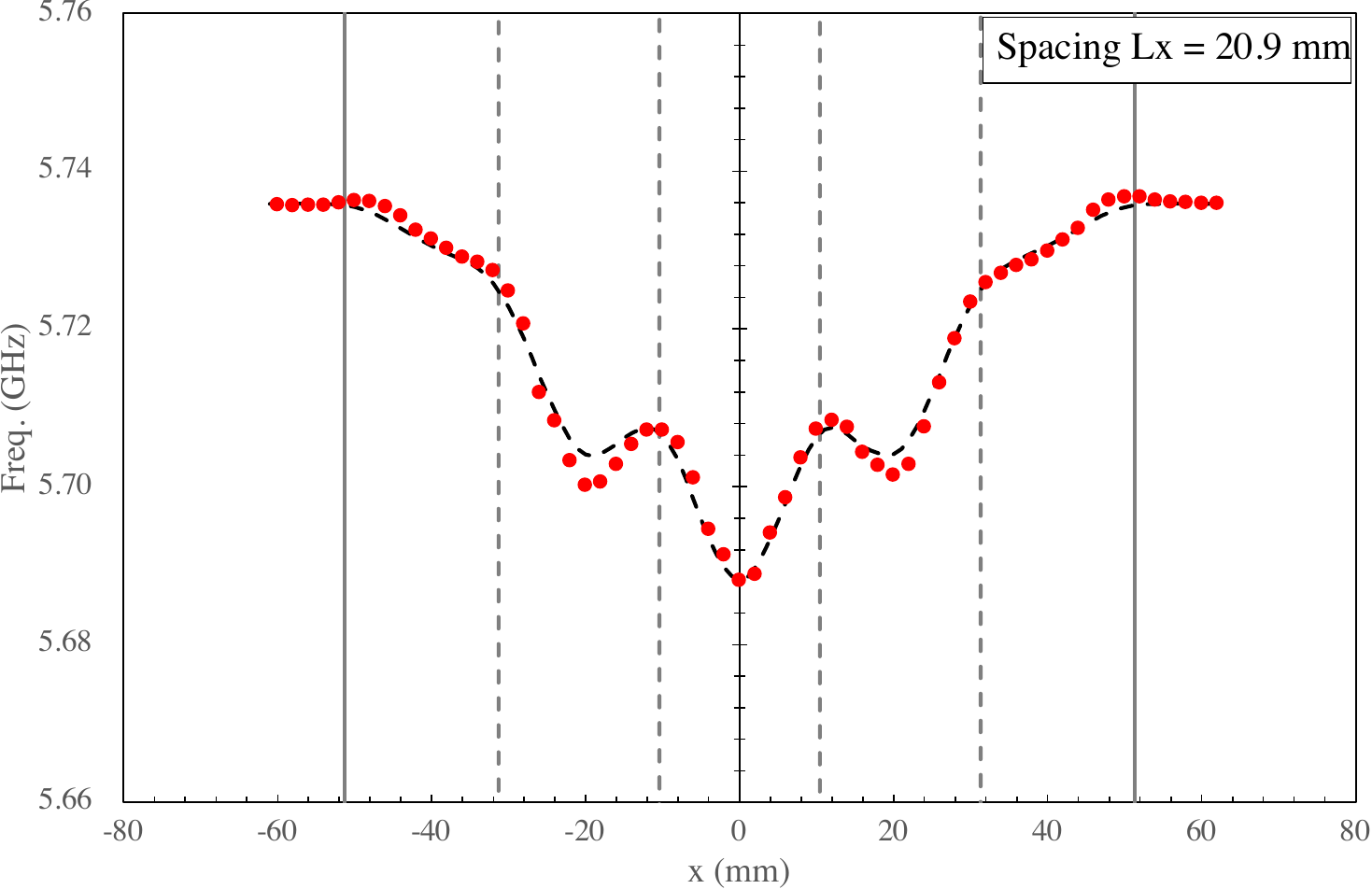}} \label{subfig:fLx21mm}
\subfloat[] [$L_x$=21.8 mm]{\includegraphics[width=1.98in, angle=0, clip]{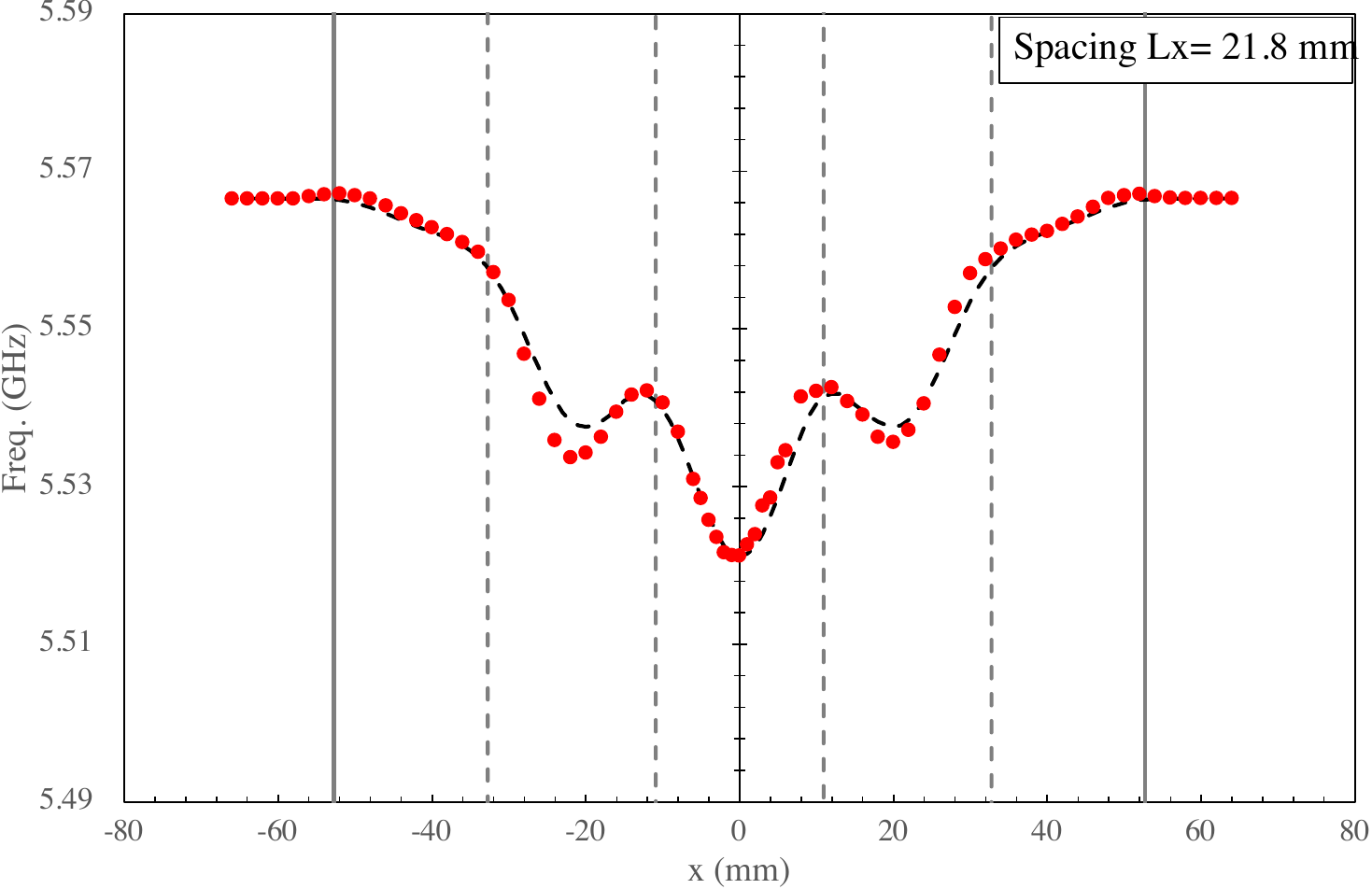}} \label{subfig:fLx22mm}
\subfloat[] [$L_x$=22.9 mm]{\includegraphics[width=1.98in, angle=0, clip]{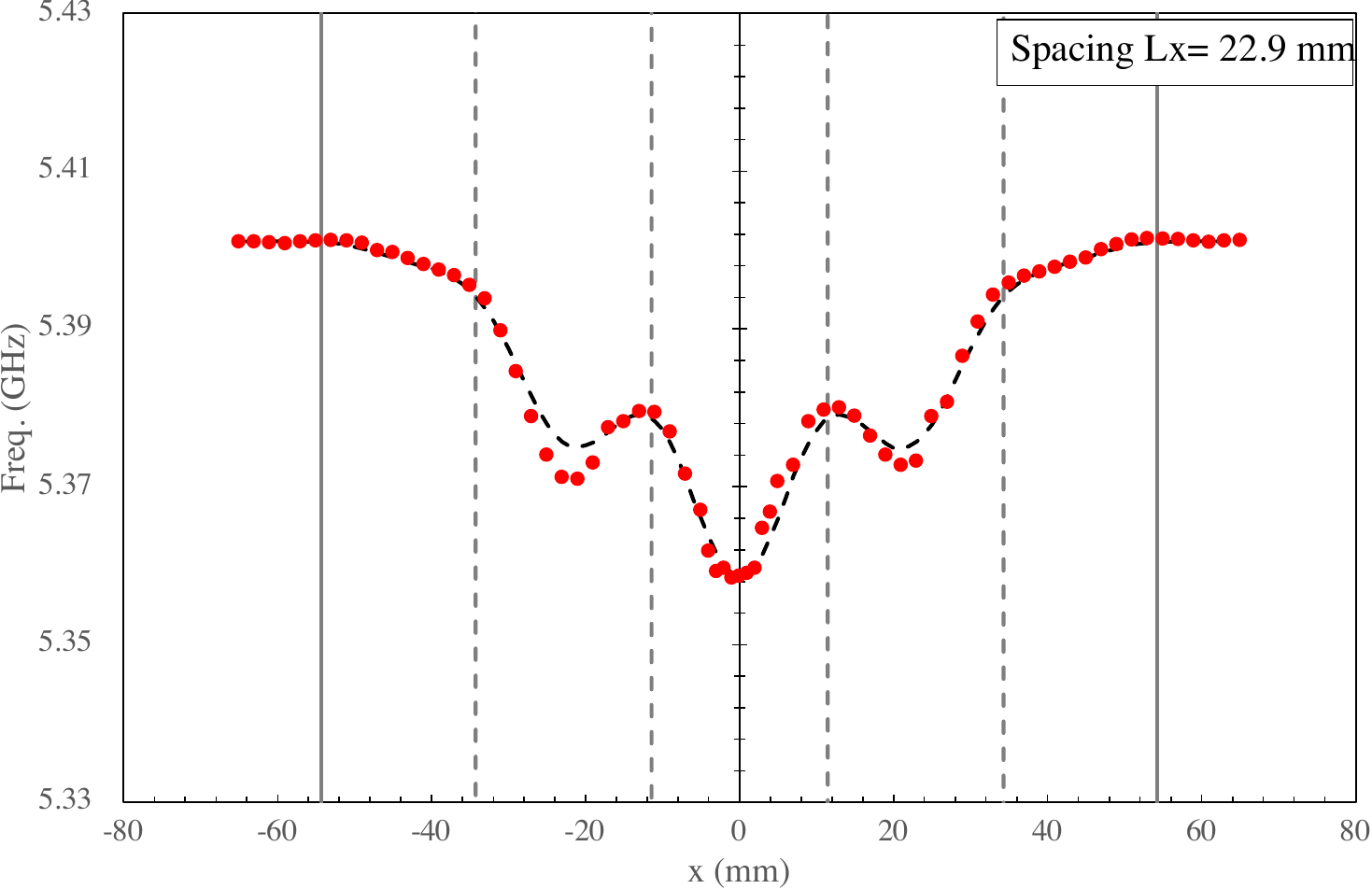}} \label{subfig:fLx23mm}

\subfloat[] [$L_x$=23.9 mm]{\i\includegraphics[width=1.98in, angle=0, clip]{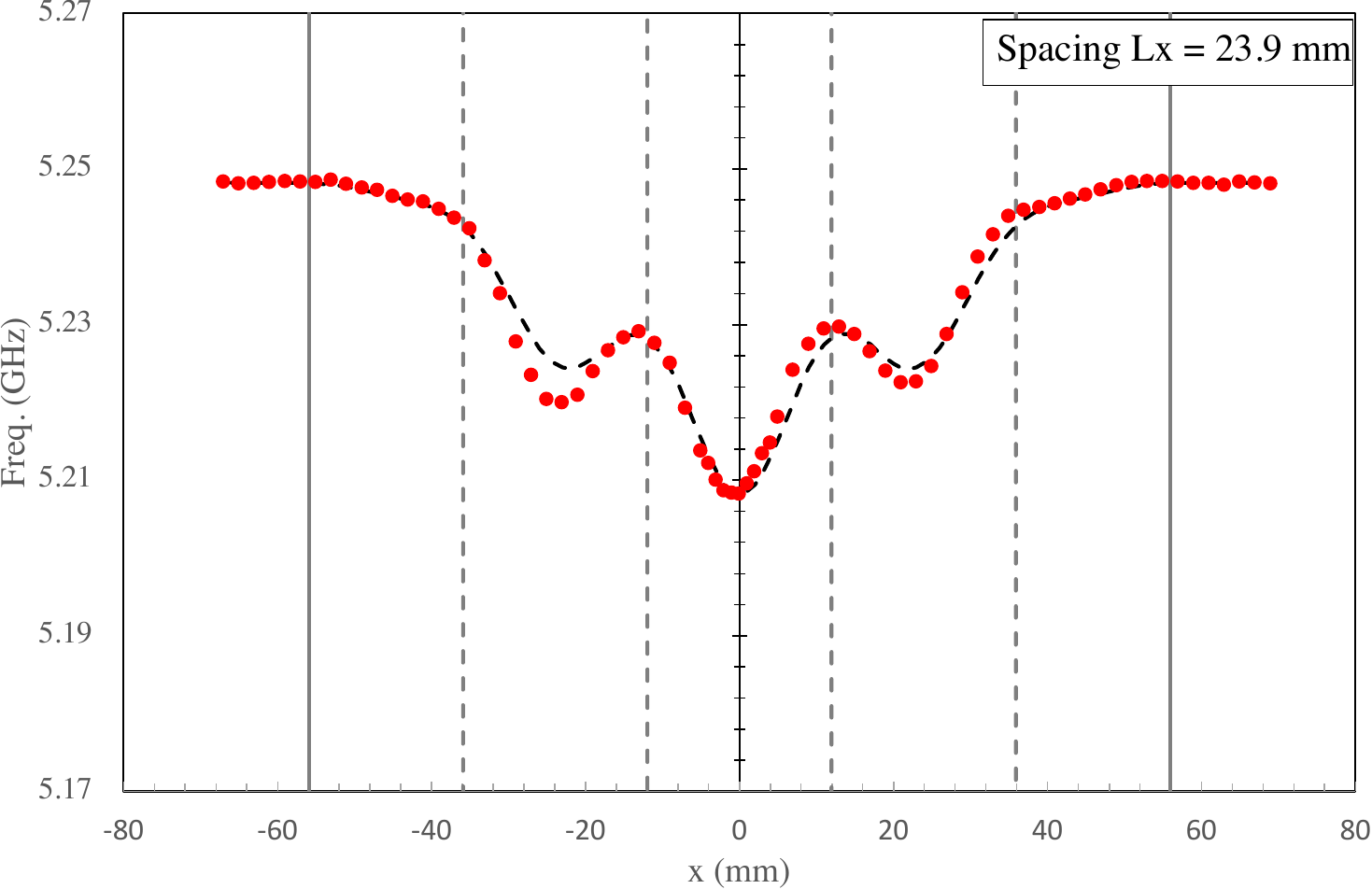}} \label{subfig:fLx24mm}
\subfloat[] [$L_x$=25.0 mm]{\includegraphics[width=1.98in, angle=0, clip]{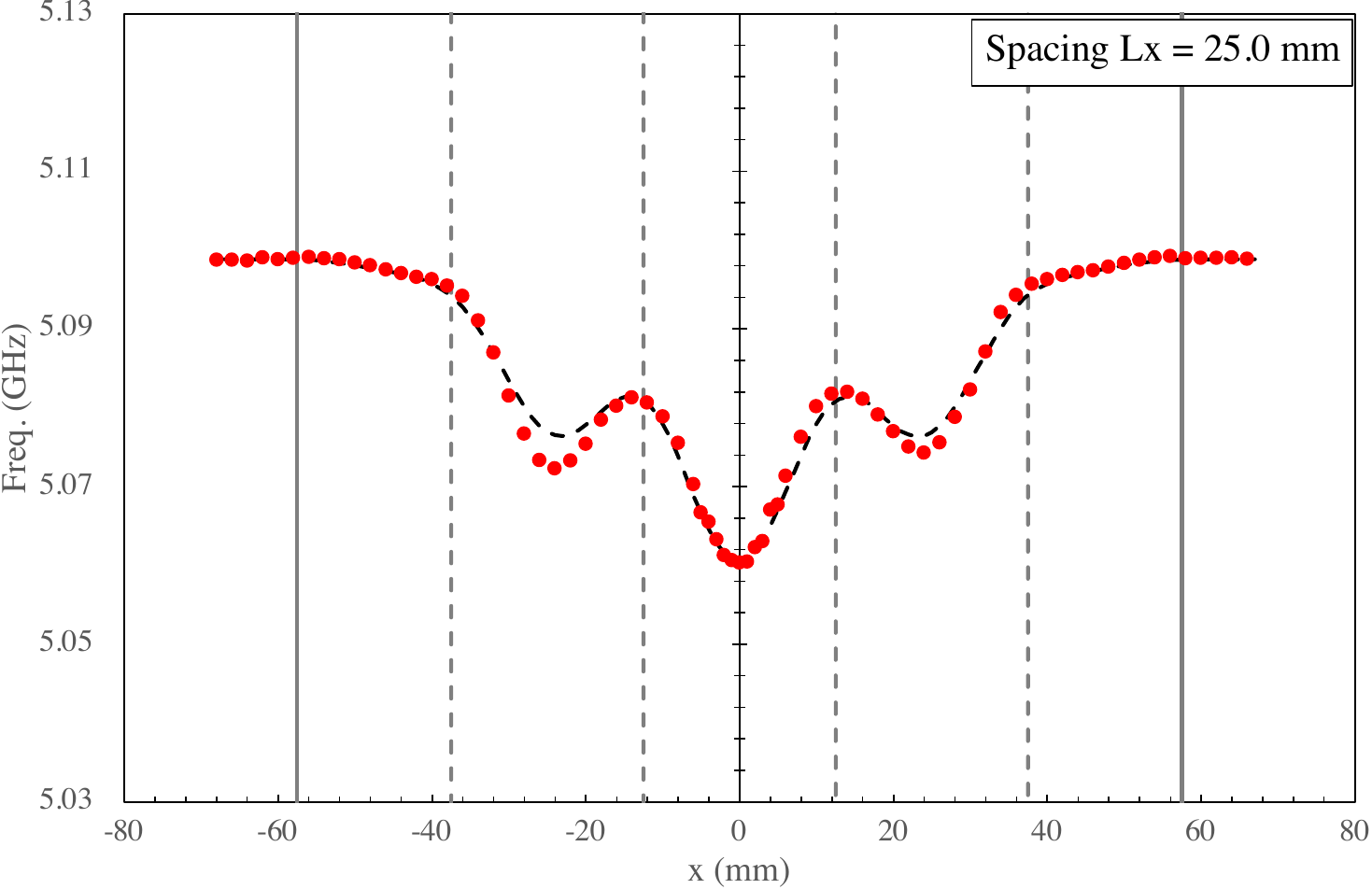}} \label{subfig:fLx25mm}

\caption{Results of the bead pull measurements in Cavity A.
From (a) to (l), $L_x$ is changed from 16 mm to 25 mm in 1 mm step. 
Red circles are experimental data and the broken lines are the estimated frequency shift by $E(x)^2$.
The frequency shifts are proportional to the squire of the electric field strength.
The measurements tell that the electric field around the grids is lower but the shape is similar to ${\rm TM_{010}}$ and no node is found in all cases.
In all plots, two thick vertical lines and thin vertical broken ones represent the positions at the cavity inner wall and the ones at the grids.}
\label{fig:eField}
\end{center}
\end{figure}

The summary of the field profile measurements is given by Fig. \ref{fig:eField}.
The experimental data and the estimated frequency shifts in each $L_x$ are also shown by closed circles and by the dotted line, respectively.
The experimental data clearly show that 
the electric fields of the developed cavity have no node and are ${\rm TM_{010}}$-like mode.
This is one of the most important results in this paper.
Furthermore, the profile of the electric field agrees well with the simulation.
Especially when $L_x$ is larger than $\simeq 20$ mm, 
it is a very good match.
These clearly indicate that the modes are very close to the simulated ones, and no mode localization is happened.
On the other hand, 
in the short range of $L_x$, 
especially in the case of $L_x \sim 17$ mm, 
the deviation is not negligible, 
and in the case of $L_x = 16 $mm, the difference is large.
These results could be interpreted as the mode localization, especially $L_x=16$ mm. 
If the electric field is not as expected one, it will become a systematic error of $G$-factor,
\begin{equation}\label{eqn:deltaG}
	\frac{\Delta G}{G}= +2\frac{\int_V dv (\delta E_z)}{\int_V dv E_z} - \frac{\int_V dv \delta (\bm  E^2)}{\int_V dv {\bm E^2}} .
\end{equation}
In the case of ${\rm TM_{010}}$-like mode, $|E| = |E_z|$, the errors of the denominator and the numerator cancel each other out in the first order approximation.
However, from a conservative standpoint, we estimate the systematic error.
The measured frequency shift on $y$-axis can be used as an estimator of the second term in Eqn. \ref{eqn:deltaG},
\begin{equation}\label{eqn:deltaGError}
	 \frac{\int_V dv \delta (\bm  E^2)}{\int_V dv {\bm E^2}}  \sim \frac{\int dx ((\Delta f)_{\rm observed}-(\Delta f)_{\rm simulation})}{\int dx (\Delta f)_{\rm simulation}}=\frac{\Sigma_x \delta( \Delta f)}{\Sigma_x \Delta f}.
\end{equation}
This value at each $L_x$ is given in Table \ref{tbl:errorEstimator}.
Even in the worst case, $L_x = 16$ mm, 
it is 9.6\%. 
From $L_x = 19.1$ mm to 25 mm, they are small enough.
It is necessary to measure the profile for actual experiments, but we can expect that it will not make a large contribution to $\Delta G$.
\begin{table}[!h]
\caption{The values of $|\frac{\Sigma \delta( \Delta f)}{\Sigma \Delta f} |$  }
\label{tbl:errorEstimator}
\centering
\begin{tabular}{|c|c|c||c|c|c|}
\hline
Setup name & $L_x$ (mm)  &  $| \frac{\Sigma \delta (\Delta f)}{\Sigma \Delta f} |$ (\%)&
	Setup name & $L_x$ (mm) &  $| \frac{\Sigma \delta (\Delta f)}{\Sigma \Delta f} |$ (\%) \\	
\hline
S-16& 15.9 &9.4& L-20& 19.8 & 1.7 \\
S-17& 16.9 &6.7&L-21& 20.9& 0.3 \\
S-18& 17.9 &7.5&L-22& 21.8& 1.4\\
S-19&19.1 &1.5&L-23& 22.9& 0.9\\
S-20& 20.0 &3.7&L-24& 23.9 & 0.2\\	
       &         &   &L-25& 25.0& 0.2\\
\hline
\end{tabular}
\end{table}

\subsection{Results of the large cavity  (180 mm $\times$ 180 mm $\times$ 20 mm $\times$ 2 )}\label{subsec:resultsOfLCavity}
\begin{figure}[!h]
\begin{center}
\centering
\subfloat[] [$S_{21}$ spectrum of the large cavity]{
{\includegraphics[width=2.7in, angle=0, clip]{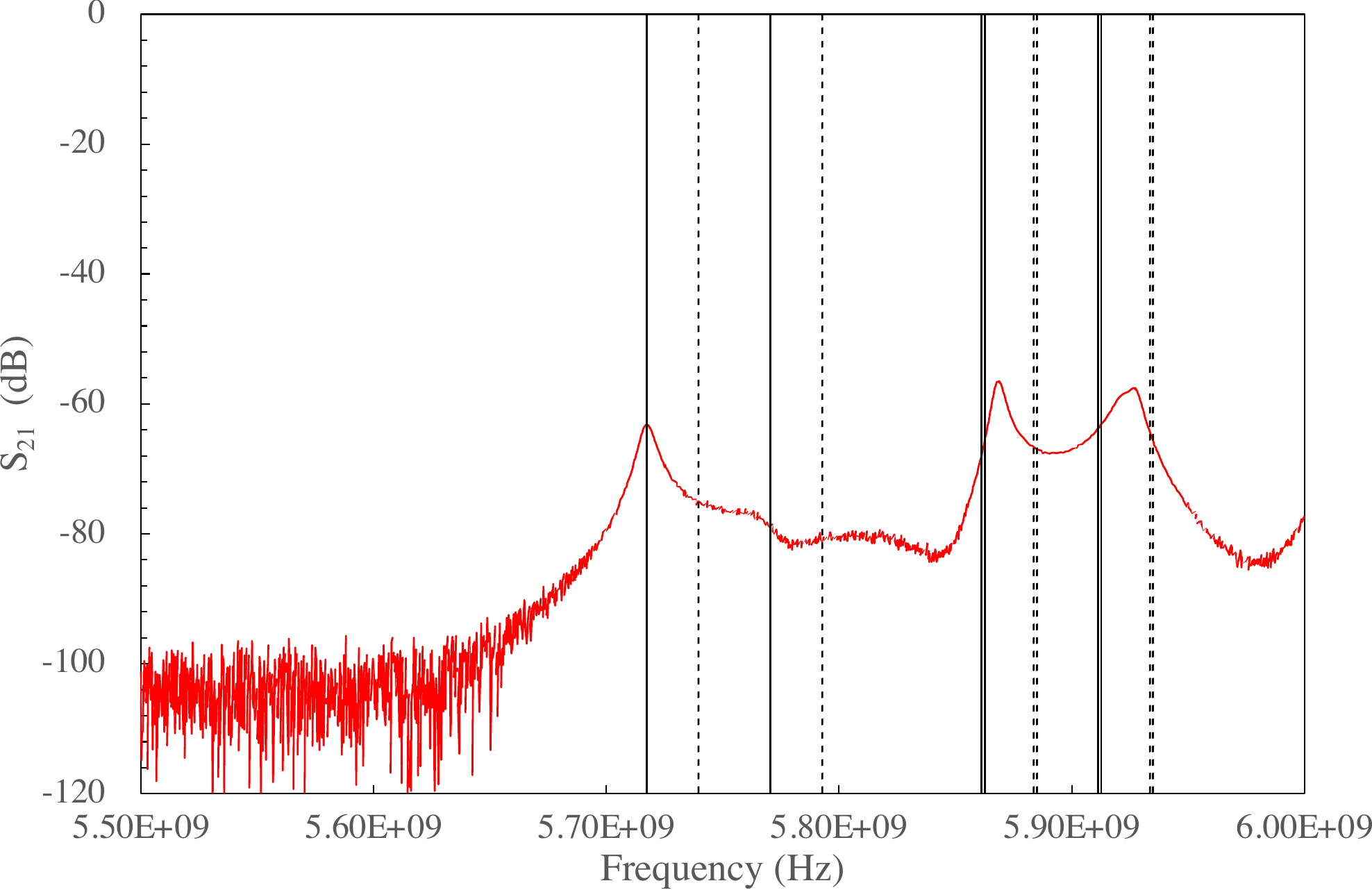}}
}\label{subfig:spectrumLCavity}
\subfloat[] [Resonance frequency shift by the bead pull method.]{
{\includegraphics[width=2.7in, angle=0, clip]{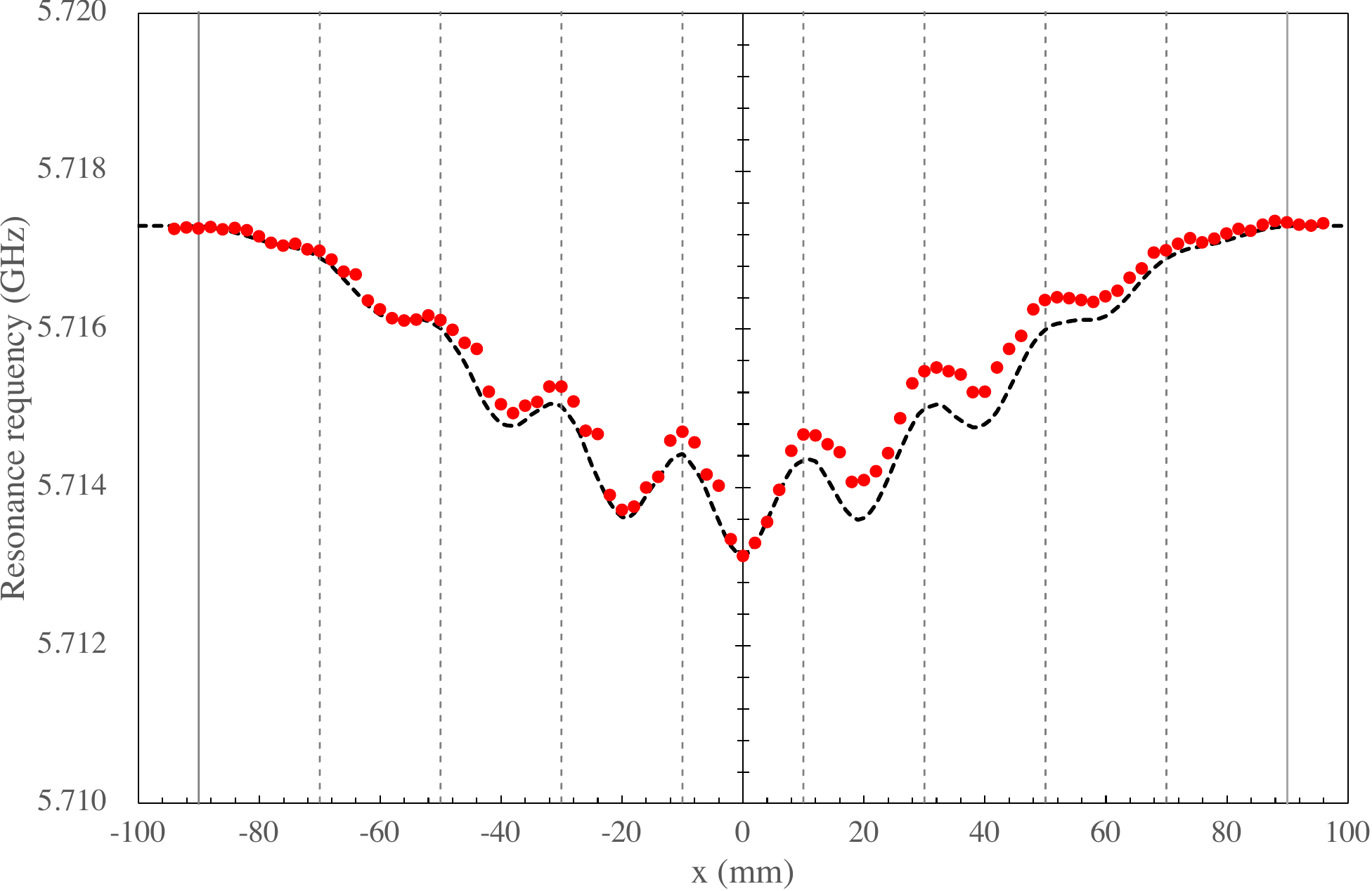}}
}\label{subFig:eFieldLCavity}
\caption{The $S_{21}$ spectrum of the large DRiPC cavity (180 mm $\times$ 180 mm $\times$ 20 mm $\times$ 2) (Left). The red line is the measured spectrum. The four dotted lines represent the simulated resonant frequencies. 
The back solid lines are the ones shifted from the dotted lines to meet the lowest-frequency experimental value. 
The right panel shows the result of the bead pull measurement on the lowest resonance frequency. The red circles are the experimental data and the black broken line is given by the simulated field values at each $x$. In this estimation, the normalization was set at $x=0$ mm. The four dotted and the two solid gray lines represent the position of the grids and the inner walls of the cavity, respectively.  }
\label{fig:largeCavity}
\end{center}
\end{figure}

Based on the results above, we then investigated the large cavity, Cavity B,  illustrated in Fig. \ref{fig:3ViewLCavity}.
We measured resonance frequencies, $Q$-value, and electric field profile
by $S_{21}$ measurements.
The experimental setup was the same as the previous measurements.
The only one difference was loop antenna positions.
Each of the two loop antennas
was set at the hole 
at $y=20$ mm in the upper and the lower box, respectively. 
The $S_{21}$ spectrum by the VNA was shown in the left panel of Fig. \ref{fig:largeCavity}.
The lowest resonance frequency was at 5.72 GHz, and the next one was not so clear but was at 5.86 GHz.
There was a deviation of $\sim 22$ MHz from the simulation result in the lowest frequency shown by the dotted line.
The black lines are the lines shifted by $\sim 22$ MHz to match the lowest mode frequencies given by the data and the simulation.
The interval between the lowest and the second peak was well matched to the simulation result, and we concluded 
the resonance at 5.72 GHz was the mode required for axion searches.
We calculated the $Q$-value to find $Q_0= 7.6\times 10^2$, 
which was a moderate number.

The electric field strength of this mode was measured by the bead pull method.
The right in Fig. \ref{fig:largeCavity}  
shows the frequency shift in the lowest resonance frequency 
as a function of the bead position 
on $x$-axis.
We also plot the estimated frequency shift given by the simulation. 
The estimation procedure was the same as Cavity A.
We confirmed ${\rm TM_{010}}$-like mode structure even in this large cavity
and the electric field is well matched to the simulation data.
So, a high $G$-factor as calculated by the simulation value, $G=0.63$, can be expected in this cavity.
If we assume a cuboid cavity with a resonance frequency of 5.72 GHz 
for ${\rm TM_{010}}$, 
the volume is $V_{\rm Cuboid}= (3.71\,\, {\rm cm})^3= 5.11 \times 10^1$ ${\rm cm^3}$, 
and we compare it to that of Cavity B, the volume ratio is 25.
These results lead us that a large DRiPC cavity could enhance axion-signal due to the large volume and high $G$-factor.

\section{Conclusion}\label{sec:conclusion}
The two DRiPC cavities made of oxygen-free copper were developed for the halo-scope experiments. 
The size of the smaller one was 100 mm in length, 100 mm in width, 
and 10 mm in height.
The larger one was a coupled cavity with a stack of 180 mm length, 180 mm width, and 20 mm height.
To cover 5 to 6 GHz, 
4 mm-diameter grids were introduced with 20 mm spacing.
The resonance frequency and $Q$-value of the smaller cavity were measured by changing the grid spacing in the $x$ direction, $L_x$. 
The $Q$-values were not very high, and it is necessary to be improved in the future. 
The measured three lowest resonance-frequencies 
showed  the excellent agreements with the ones simulated by the finite element method.
This result strongly suggests that the lowest-order resonance mode is a ${\rm TM_{010}}$-like mode. 
The tuning range of the lowest-order resonance frequency was quite wide, achieving 
a total of 27.7\% around 5.87 GHz without mode crossing. 
In the range of $L_x = 17$ -- 25 mm, the frequency change was linear, and the amount of change per unit length was 164 MHz/mm.
With the bead pull method, 
the distribution of the electric field intensity in the cavity in the lowest resonance mode was measured 
and it was confirmed 
that  it is  ${\rm TM_{010}}$-like.
The resonance frequency shifts were in accordance with the simulated ones.
We also evaluated the systematic error of $G$-factor in the DRiPC cavity, 
and it was found to be small enough to rely on the $G$-factor obtained by the simulation in future experiments. 
The ${\rm TM_{010}}$ mode was also observed in the large size coupled cavity. 
These results demonstrate the possibility to create a resonant cavity with the three essential elements 
for axion searches:
a large volume, high $G$-factor, a wide frequency-range tunability.

\section*{Acknowledgment}
We thank Prof. R. Kawabe (NAOJ), Prof. H.~Ogawa (Osaka Prefecture University), and Dr. Y.~Hasegawa (JAXA) for their useful discussions and advice at the early stage of this study.
This work was supported by the Japan Society for the Promotion of Science (JSPS) KAKENHI Grants Grant-in-Aid for Scientific Research on Innovative Areas No. 17H02883 and 
Grant-in-Aid for Scientific Research (B) No.19H05809.

%

\vspace{0.2cm}
\noindent

\let\doi\relax


\end{document}